\documentclass[8pt]{article}

\usepackage{
graphicx,amsmath,amssymb,bm,nccmath}

\newtheorem{prop}{Prop}[section]
\newtheorem{lemma}[prop]{Lemma}

\newtheorem{theorem}[prop]{Theorem}

\title{
Expansion for quantum perturbations in random  spin systems 
}
\author{C. Itoi$^1$, K. Horie$^1$, H. Shimajiri$^1$ 
and
Y. Sakamoto$^2$,\\
 $^1$Department of Physics, GS $\&$ CST, Nihon University, \\
Kandasurugadai, Chiyoda,  Tokyo 101-8308, Japan\\
$^2$ Laboratory of Physics, CST, Nihon University, \\
Narashinodai, Funabashi-city, Chiba 274-8501, Japan} 
\begin{document}
\maketitle


\maketitle 

\begin{abstract} Energy eigenstates in the random transverse field Edwards-Anderson (EA) model and 
the random bond quantum Heisenberg XYZ model in a $d$-dimensional finite cubic lattice
are obtained for sufficiently weak interactions.  
The Datta-Kennedy-Kirkwood-Thomas convergent perturbative expansion using the contraction mapping theorem
is developed  for quantum spin systems with  site- and bond-dependent interactions.
This expansion  enables us to obtain energy eigenstates in
the  random transverse field free spin model  perturbed by sufficiently weak longitudinal exchange interactions.
This expansion is  useful also for the EA model perturbed by sufficiently  weak  transverse fields
and bond-dependent XY exchange interactions.   In these models, their perturbations split   
the two fold degenerate energy eigenvalues  because of the
${\mathbb Z}_2$ symmetry in the unperturbed EA model.
It is shown that the energy gap between split energy eigenvalues  is exponentially small in the system size.
We provide a sufficient condition on the perturbation 
for absence of  level crossing between arbitrary energy eigenstates. \end{abstract}

%
%
%
%

\section{Introduction}
Quantum spin systems with random interactions  have been studied extensively.
Many physicists, mathematicians and computer scientists have studied these systems including
the transverse field Ising model with random interactions, since D-Wave Systems actually 
devised and produced a quantum annealer based on  these models \cite{J}.  Quantum annealer  is an optimization hardware, which is
theoretically proposed  by Finnila-Gomez-Sebenik-Stenson-Dol \cite{FGSSD} and Kadowaki-Nishimori  \cite{KN} on the basis of 
the adiabatic theorem in quantum mechanics.
The adiabatic theorem claims that the time developed state from the ground state 
is preserved as the corresponding  instantaneous ground state 
by an infinitely slow time dependent perturbation, if the perturbation  gives no
 level crossing between the ground sate and an excited state.  
Absence of level crossing in the ground state is necessary for the precise solution obtained by 
  the quantum annealer for the optimization problem 
of the Hamiltonian in the Ising model with random  interactions. 
 Apart from problems in quantum annealing,  generally it is great concern whether or not,  quantum perturbations give
a level crossing of the energy eigenvalues in  spin modes with random interactions.

 In the present paper,
 $S=1/2$ quantum spin systems with site- and bond-dependent  interactions are studied
  in  the Kirkwood-Thomas 
  convergent perturbative expansion \cite{KT} developed by  Datta and Kennedy \cite{DK1,DK2}.
   Quantum spin systems are regarded as classical Ising systems  with quantum perturbations.
  The perturbative expansion is performed around the classical Ising systems  with diagonalized Hamiltonians.
  Some sufficient conditions on quantum interactions for the absence of level crossing is obtained  
  by a convergent perturbative expansion for sufficiently weak quantum interactions.
This expansion  enables us to obtain 
 energy eigenstates in the model for  weak quantum perturbations.  Datta and Kennedy develop the Kirkwood-Thomas expansion method and study uniform transverse field Ising model \cite{DK1} and the  Heisenberg  XXZ model \cite{DK2}.
  The expansion method given by Kirkwood and Thomas can be applied only to a restricted class of  systems whose Hamiltonians 
satisfy the Perron-Frobenius condition. Datta and Kennedy have improved the Kirkwood-Thomas  method by removing this condition.   
Banach's fixed point theorem for an arbitrary contraction mapping is used 
for functions which define  quantum spin states.   \\

Consider $d$-dimensional hyper cubic lattice  $\Lambda_L= {\mathbb Z}^d \cap (-L/2,L/2]^d$ 
with an even integer $L >0$.  Note that the lattice $\Lambda_L$  contains  $ L^d$ sites.
$S_\Lambda:=2^{\Lambda_L}$ denotes the collection of sub-lattices of $\Lambda_L.$
Each lattice site $i\in \Lambda_L$ has  an operator-valued spin vector
 $(\sigma_i^x, \sigma_i^y, \sigma_i^z )$ defined by the Pauli matrices.
 For a sub-lattice $X \in S_\Lambda$,  denote
 $$
 \sigma_X^w:= \prod_{i \in X} \sigma_i^w
 $$
 for $w=x,y,z$.
 Define a set of nearest neighbor bonds by
 $$B_\Lambda = \{ \{i, j\} | i,j \in   \Lambda _L,  |i-j|=1\}.$$ Note $|B_{\Lambda}|=|\Lambda_L| d.$
 A bond spin  $\sigma_b$ denotes   
$\sigma_b^a=\sigma_i^a \sigma_j^a $
 for a bond $b= \{i,j\} \in B_{\Lambda}$ and $w=x,y,z$.
 Let $ \bm h :=(h_i)_{i \in \Lambda_L} $, 
 $\bm J:=(J_b)_{b \in B_{\Lambda}}$ and $ \bm \epsilon :=(\epsilon_b^x, \epsilon_b^y)_{b \in B_\Lambda} $ be  sequences of arbitrary real numbers.
Although these numbers can be random variables, here we do not assume their specific distributions.
Define the following  functions of a sequence of 
spin operators $\bm \sigma:=(\sigma_i^w)_{i\in \Lambda_L,w=x,y,z}$ and sequences of coupling constants 
$\bm h, \bm J, \bm \epsilon$
\begin{eqnarray}
&&H_\Lambda^{\rm x}(\bm \sigma, \bm h) :=-\sum_{i\in \Lambda_L}h_i  \sigma_i^x,   \label{x}\\
&&H_\Lambda^{Z}(\bm \sigma,  \bm J):= - \sum_{b \in B_{\Lambda}} J_b\sigma_b ^z
\label{Z}\\
&&H_\Lambda^{\rm XY}(\bm \sigma, \bm \epsilon):= - \sum_{b \in B_{\Lambda}}( \epsilon_b^x \sigma_b^x+
  \epsilon_b^y \sigma_b^y).
\label{XY} 
\end{eqnarray}
In the present paper,  we study the random transverse field Edwards-Anderson model
 \begin{equation}
 H_\Lambda^{\rm xZ} (\bm \sigma, \bm h, \bm J) := H_\Lambda^{\rm x}(\bm \sigma, \bm h)+H_\Lambda^{Z}(\bm \sigma,  \bm J),
 \label{xZ}
 \end{equation}
 and the random bond Heisenberg XYZ model
 \begin{equation}
 H_\Lambda^{\rm XYZ} (\bm \sigma, \bm J, \bm \epsilon) := H_\Lambda^{Z}(\bm \sigma, \bm J)+H_\Lambda^{\rm XY}(\bm \sigma,  \bm \epsilon),
 \label{XYZ}
 \end{equation}
 in a convergent expansion.  

 First in the present paper,  the random  transverse field Edwards-Anderson (EA) model defined by the Hamiltonian (\ref{xZ})  is studied
 in the expansion around unperturbed model $\bm J =\bm 0$. If 
 longitudinal exchange interactions  
 are switched off, then this model  defined by the Hamiltonian
  $H_\Lambda^{\rm xZ}(\bm \sigma, \bm h, \bm 0)$ becomes a free spin model under random transverse fields.  
 All eigenstates and energy eigenvalues are obtained trivially in the free spin model, and it can be shown that 
 the energy eigenstates in this free spin model are not degenerate for almost all transverse fields $\bm h$. 
 The weak longitudinal exchange interactions can be treated by a simple perturbative expansion.  
 
 Next,  the random  transverse field EA model defined by the Hamiltonian (\ref{xZ}) is studied
 in the expansion around $\bm h =\bm 0$.
 Although the unperturbed Hamiltonian $H_\Lambda^{\rm xZ} (\bm \sigma, \bm 0, \bm J)$ is also diagonalized,
all energy eigenvalues  are two fold degenerate because of the  $\mathbb Z_2$ symmetry of the  global spin flip \cite{EA}.  
For sufficiently  weak transverse  fields, a perturbative expansion around $\bm h=\bm 0$ 
can be done  for almost all longitudinal exchange interactions $\bm J$.  The degenerate  unperturbed energy eigenvalues  
are split by the perturbation of transverse fields. It is proven that the energy gap between these 
 split energy eigenvalues is exponentially small in the system size.  It is proven also that  the sufficiently weak  perturbation cannot produce  the level crossing under a condition to break the $\mathbb Z_2$ symmetry.
 
Finally, the random bond Heisenberg XYZ model  defined by the Hamiltonian (\ref{XYZ}) is studied  in the perturbative expansion
around $\bm \epsilon =\bm 0$.
This model is regarded as the EA model 
$ H_\Lambda^{\rm XYZ} (\bm \sigma, \bm J, \bm 0)$ perturbed  by the XY-exchange interactions.
   We obtain all energy eigenstates  in random bond quantum  Heisenberg model, if its XY  exchange interactions are  sufficiently weak.
  Similar properties to those in weak transverse field EA model are obtained also in this model because of the $ \mathbb Z_2$ symmetry.
  As in the weak transverse field EA model, the weak XY interactions split the unperturbed degenerate energy eigenvalues, and
the energy gap between these  is also exponentially small in the system size. 
It is proven also that the weak XY interactions cannot produce the level crossing under a condition to break  the $\mathbb Z_2$  symmetry. 

The present paper is organized as follows. In section two, an arbitrary 
 energy eigenstate in the transverse field EA model defined by the Hamiltonian  (\ref{xZ})
  is obtained in the Datta-Kenedy-Kirkwood-Thomas expansion for the perturbation  (\ref{Z})
  with sufficiently weak longitudinal exchange interactions. 
In section three,  this model is studied in another expansion for perturbation  (\ref{x}) with weak transverse fields. 
In section four, the random bond Heisenberg XYZ model defined by (\ref{XYZ})
 is studied in the expansion for perturbation (\ref{XY}) with weak XY-exchange interactions.

\section{Transverse field EA model  around the free spin model}

In this section, we study the model defined by the Hamiltonian (\ref{xZ}) around $\bm J=\bm 0$ in the Datta-Kennedy-Kirkwood-Thomas expansion.
\subsection{Unperturbed model}
The following lemma guarantees  the non-degenerate energy eigenstates 
in  the  random transverse field free spin model  
defined by the Hamiltonian (\ref{x}). \\

\noindent
{\lemma  \label{L1} Consider  the unperturbed model  defined by the Hamiltonian (\ref{x}) 
at $\bm J=\bm 0$ in $d$-dimensional hyper cubic lattice $\Lambda_L$.
The Hamiltonian takes different values 
$$
H_\Lambda^{\rm x}(\sigma,\bm h)  \neq H_\Lambda^{\rm x}(\sigma',\bm h) ,
$$
for  any two different spin configurations $\sigma, \sigma'\in \{1,-1 \}^ {\Lambda_L} $ for almost all $\bm h \in {\mathbb R}^{\Lambda}$}.\\

\noindent
{\bf Proof. }
Let  $i_1, i_2, i_3, \cdots  \in \Lambda_L$ be a sequence of  different
 sites.  Define a sub-lattice  $X_N ( \in S_\Lambda) $  by
$$
X_N:=   \bigcup_{n=1}^N \{ i_n \} .
$$
The following mathematical inductivity with respect to $N$ enables us to prove this lemma.\\
For $N=1$,   $X_1=\{ i_1\}$  is a single bond. 
 $$
 H_{X_1} ^{\rm x}(\sigma, \bm h) = -h_{i_1} \sigma_{i_1}.
 $$
 Since the Hamiltonian takes different values 
 for $\sigma_{i_1} =\pm1$,  this lemma is valid for $N=1$ and $h_{i_1} \neq 0$.\\ 
 For an arbitrary positive integer $N$, assume the validity of this lemma. Then, 
 \begin{equation}
H_{X_N}^{\rm x}(\sigma,\bm h)  \neq H_{X_N}^{\rm x}(\sigma',\bm h),
\label{Nh}
\end{equation}
is valid for any two different configurations $\sigma, \sigma' \in \{1,-1\}^{ X_{N}}$ for almost all $\bm h$. \\
For $N+1$, 
 let $\sigma, \sigma'\in \{1,-1\}^{ X_{N+1}}$ be two different configurations.
Consider
the  equation  for $h_{i_{N+1}}$
\begin{equation}
H_{X_{N+1}}^{\rm x}(\sigma,\bm h)   = H_{X_{N+1}}^{\rm x}(\sigma',\bm h), 
\label{eqh}
\end{equation}
which has the following representation in terms of $H_{X_N}^x$ and $h_{i_{N+1}}$
$$
H_{X_N}^{\rm x}(\sigma|_{X_N},\bm h)   - h_{i_{N+1}} \sigma_{i_{N+1}} = H_{X_N}^{\rm x}(\sigma'|_{X_N},\bm h)   - h_{i_{N+1}} \sigma'_{i_{N+1}}.
$$
Since 
$|X_{N+1}\setminus X_N| = 1$,   $\sigma_{i_{N+1}} =\sigma'_{i_{N+1}}$ 
implies 
the assumption (\ref{Nh}).  Then, 
the equation (\ref{eqh}) has no solution for $\sigma_{i_{N+1}} =\sigma'_{b_{N+1}}$. For $ \sigma_{i_{N+1}} -\sigma'_{i_{N+1}} = \pm 2$,
 the corresponding solutions of the equation (\ref{eqh}) are given by
\begin{equation}
h_{i_{N+1}} =\pm  \frac{1}{2}[H_{X_N}^{\rm x}(\sigma'|_{X_N},\bm h) - H_{X_N}^{\rm x}(\sigma|_{X_N},\bm h)  ].
\label{solh}
\end{equation}
Therefore, 
 $$
H_{X_{N+1}}^{\rm x}(\sigma,\bm h)   \neq  H_{X_{N+1}}^{\rm x}(\sigma',\bm h) , 
$$ 
is valid also for  $N+1$  for almost all $h_{i_{N+1}} \in {\mathbb R}$ except  the solutions (\ref{solh}).  Then, this lemma is valid for an arbitrary positive integer $N$.
This completes the proof. $\Box$

\subsection{Reference state}
Define
$$\sigma_X := \prod_{i\in X} \sigma_i,$$
and  for $X=\phi$, $\sigma_\phi:=1$.
$\Sigma_\Lambda :=\{1,-1 \}^{\Lambda_L}$ denotes the collection of sequences of eigenvalues of $\sigma_i^z$.
 Note the following identity  for $X, Y \subset \Lambda_L$
\begin{equation}
2^{-|\Lambda_L|} \sum_{\sigma\in \Sigma_\Lambda} \sigma_X\sigma_Y = I(X=Y) =: \delta_{X,Y},
 \label{on}
\end{equation}
where  
an indicator $I(e)$ for an arbitrary event  $e$ is defined  by  $I({\rm true})=1$ and $I({\rm false}) =0$. 
Here, we discuss a convergent  expansion around $\bm J= \bm 0$.
In the case for $\bm J= \bm 0$,
 the  model  defined by the Hamiltonian (\ref{xZ}) becomes the free spin model, and Lemma \ref{L1} guarantees 
 absence of degeneracy in energy levels.  
   For an arbitrary sub-lattice $D \in S_\Lambda$,  define a sequence $s^D\in \{1,-1 \}^{ \Lambda_L } $ by
\begin{equation}
s^D_i=
\left\{
\begin{array}{lll}
-1 & ( \ i \in D \ ) \\
1 & ( \  i \notin D \ ) 
\end{array}
\right.
\label{sD}
\end{equation}
The following  state satisfies
\begin{eqnarray}
-\sum_{i\in \Lambda_L}h_i \sigma^x_i  \sum_{\sigma \in \Sigma_\Lambda}\sigma_D | \sigma \rangle= 
-\sum_{i\in \Lambda_L}h_i s_i^D \sum_{\sigma \in \Sigma_\Lambda}\sigma_D | \sigma \rangle.
\label{bondexp} 
\end{eqnarray}   
This state  gives each state corresponding to $D \in S_\Lambda$ for $\bm J=\bm 0$.

Let $\psi(\sigma)$ be a function $\psi: \{-1,1\}^{\Lambda_L} \to {\mathbb  R }$ of a sequence,
and express the energy  eigenstate of the Hamiltonian (\ref{xZ}) for weak exchange interactions
$$
|D \rangle= 2^{-|\Lambda_L|/2}\sum_{\sigma \in \Sigma_\Lambda}\sigma_D \psi(\sigma)| \sigma \rangle.
$$ 
The normalization $\langle D|D\rangle=1$ requires
\begin{equation}
\sum_{\sigma\in \Sigma_\Lambda} \psi(\sigma)^2=2^{|\Lambda_L|}.
\label{normalization1}
\end{equation}
Note that  $\psi(\sigma) =1$
for $J_b=0$  is given by  the state corresponding to $D \in S_\Lambda$.
 This state $|D\rangle$ defined by $D:=\{i\in \Lambda_L|  h_i < 0 \} $ becomes  the ground state  for $J_b=0$, and 
$s_i^D= h_i/|h_i|.$ 
The eigenvalue equation defined by
$$
 H_\Lambda^{\rm xZ}(\sigma, {\bm h},\bm J) |D \rangle =E_D |D \rangle
$$
 is written in 
$$
-( \sum_{b\in B_\Lambda} J_b \sigma_b^z +\sum_{i\in \Lambda_L} h_i \sigma_i^x ) |D \rangle = E_D |D\rangle.
$$
Using  $\sigma_i ^x |\sigma \rangle = | \sigma^{(i)} \rangle$ and 
$
\sigma_b^z| \sigma\rangle
=\sigma_b  |\sigma \rangle$, 
the eigenvalue equation can be represented in terms of $\psi(\sigma)$.
\begin{equation}
 \sum _{b\in B_\Lambda} J_b\sigma_{b} \sigma_D\psi(\sigma)  + \sum_{i\in \Lambda_L}h_i  \sigma_D^{(i)} \psi(\sigma^{(i)}) 
 = -E_D\sigma_D \psi(\sigma),
\label{eigeneqpsi1}
\end{equation}
where $\sigma^{(i)}$  denotes a  spin configuration replaced  by $\sigma_i  \to -\sigma_i$.
Therefore
\begin{equation}
 \sum _{b\in B_\Lambda} J_b  \sigma_{b} + \sum_{i\in \Lambda_L} h_i \frac{\sigma_D^{(i)}  \psi(\sigma^{(i)})}{\sigma_D\psi(\sigma)} 
= -E_D.
\label{eigeneqpsi}
\end{equation}
To obtain the Kirkwood-Thomas equation for the state, represent the function $\psi(\sigma)$ in terms of 
 a real valued  function $g(X)$ of an arbitrary sub-lattice $X \in S_\Lambda $,
\begin{equation}
\psi(\sigma) = \exp \Big[-\frac{1}{2} \sum_{X \in S_\Lambda } g(X) \sigma_X \Big].
\end{equation}
Note the following relations 
\begin{equation}
\psi(\sigma^{(i)}) = \exp \Big[ -\frac{1}{2}\sum_{X \in S_\Lambda } g(X) \sigma_X +\sum_{X \in S_\Lambda }  I(i \in X)g(X) \sigma_X  \Big],
\end{equation}
then
$$
 \frac{\psi(\sigma^{(i)})}{\psi(\sigma)}  =  \exp \Big[ \sum_{X \in S_\Lambda }  I(i \in X)g(X) \sigma_X  \Big].
$$
Note also 
$$
\frac{\sigma^{(i)}_D}{\sigma_D}= s_i^D.
$$
These and the eigenvalue equation (\ref{eigeneqpsi1}) give 
\begin{equation}
 \sum _{b\in B_\Lambda} J_b\sigma_b
 + \sum_{i\in \Lambda_L}h_is_i^D   \exp \Big[\sum_{X \in S_\Lambda }  I(i \in  X)g(X) \sigma_X \Big]  = -E_D.
\end{equation}
We expand the exponential function in power series. 
\begin{eqnarray}
 &&\sum _{b\in B_\Lambda} J_b\sigma_b
  +E_D+ \sum_{i\in \Lambda_L}h_is_i^D +  \sum_{X \in S_\Lambda }   \sum_{i\in \Lambda_L}h_is_i^D I(i \in X) g(X)  \sigma_X\nonumber \\
 &&+\sum _{i \in \Lambda_L}h_is_i^D\exp^{(2)} \Big[ \sum_{X \in S_\Lambda }  I(i \in X)g(X) \sigma_X \Big]=0,
\end{eqnarray}
The orthonormalization property (\ref{on})  gives
\begin{equation}
E_D =-\sum _{i\in \Lambda_L} h_is_i^D  -\sum _{i \in \Lambda_L} h_is_i^D \sum_{k=2}^\infty \frac{1}{k!}  \sum_{ X_1, \cdots, X_k\in S_\Lambda}
\delta_{
X_1 \triangle \cdots \triangle X_k, \phi} \prod_{l=1}^k g(X_l)I(i \in X_l),
\end{equation}
and  for $X\neq \phi$
\begin{eqnarray}
g(X)&=&\frac{-1}{\sum _{i\in  X} h_is_i^D} \left[\sum _{j \in \Lambda_L} h_j s_j^D \sum_{k=2}^\infty \frac{1}{k!}  \sum_{ X_1, \cdots, X_k \in S_\Lambda}
\delta_{
X_1 \triangle \cdots \triangle X_k, X} \prod_{l=1}^k g(X_l) I(j \in  X_l)\right.\nonumber\\
&&\left.+\sum_{b \in B_\Lambda} J_b \delta _{X, \{ b\}} \right] 
=: F(g)(X),
\end{eqnarray}
where $X \triangle Y:= (X \cup Y) \setminus (X\cap Y) $ for arbitrary sets $X, Y$,  and we have used $\sigma_X \sigma_Y=\sigma_{X\triangle Y}.$ The normalization (\ref{normalization1}) fixes $g(\phi)$.
The first term in the energy  eigenvalue
 is identical to that of the eigenvalue configuration for $\bm J=\bm 0$,
 and  the excited energy  of a spin configuration $\sigma$ for $\bm J=\bm 0$
is $2 \sum_{i \in  X} h_is_i^D $, where
 $X:=\{ i \in \Lambda_L| \sigma_i \neq s_i^D\}$. 
To prove uniqueness of the function $g$ for the energy eigenstate $|D\rangle$ in the transverse field EA model with a given $\bm h$
for sufficiently weak $\bm J$, define a norm for the function $g(X)$ by 
\begin{equation}
\| g \| := \sup_{i \in \Lambda_L}\sum_{X \in S_\Lambda } I(i \in  X) \Big| \sum_{j \in X}h_js_j^D \Big|  |g(X)|
,
\label{norm}
\end{equation}
Then, the following theorem can be proven.\\

{\theorem  \label{T1} Consider the transverse field EA model
defined by the Hamiltonian (\ref{xZ}).  For sufficiently weak exchange coupling constants $\bm J$ and for almost all $\bm h
\in {\mathbb R}^{\Lambda_L} $,
there is a unique  energy eigenstate, which
 corresponds to the unperturbed energy eigenstate $\sum_{\sigma \in \Sigma_\Lambda} \sigma_D|\sigma\rangle$. 
 }\\

The following lemma and the contraction mapping theorem enable us to prove this theorem.

\noindent
{\lemma  \label{L2}
Define a set $\partial X$ by
$
\partial X :=\{ \{i,j\} \in B_\Lambda | i\in X, j\notin X
\}.
$
There exists a sufficiently small constant $\delta > 0$
such that   if $\sup_{j \in\Lambda_L} \sum_{b \in \partial \{j\}} |J_b| < \frac{\delta}{2} $,
\begin{equation}
\|F(g)-F(g') \| \leq   \| g-g' \|/2,   \ \  \| F(g) \| \leq \delta, \ \ {\rm for}\  \| g\|, \| g'\| \leq \delta,
\end{equation} 
 for almost all $\bm h \in {\mathbb R}^{\Lambda_L}$.
 }\\

\noindent
{\bf Proof.} For lighter notations,  define $\triangle_k := X_1 \triangle \cdots \triangle X_k$. 
  The norm $\| F(g)-F(g') \|$ is represented in
\begin{eqnarray}
&&\| F(g)-F(g')\| \\
&&= \sup_{j \in \Lambda_L}\sum_{X\in S_\Lambda }I(j \in X)
\Big|\sum_{i \in \Lambda_L} h_is_i^D 
\sum_{k=2}^\infty \frac{1}{k!}  \sum_{X_1, \cdots, X_k \in S_\Lambda}
\delta_{\triangle_k, X}
[\prod_{l=1}^k g(X_l) - \prod_{l=1}^k g'(X_l) ]
\prod_{l=1}^kI(i \in  X_l) 
\Big|\nonumber\\
&&= \sup_{j \in \Lambda_L}\sum_{X\in S_\Lambda }\Big|
\sum_{k=2}^\infty \frac{1}{k!}  \sum_{X_1, \cdots, X_k \in S_\Lambda}I(j \in \triangle_k)
\delta_{\triangle_k, X} \sum_{i \in \Lambda_L} h_is_i^D \prod_{l=1}^kI(i \in  X_l)
[\prod_{l=1}^k g(X_l) - \prod_{l=1}^k g'(X_l) ]\Big|
\nonumber\\
&&\leq \sup_{j \in \Lambda_L}\sum_{X\in S_\Lambda }
\sum_{k=2}^\infty \frac{1}{k!}  \sum_{X_1, \cdots, X_k \in S_\Lambda}I(j \in \triangle_k)
\delta_{\triangle_k, X} \Big|\sum_{i \in \Lambda_L} h_is_i^D \prod_{l=1}^kI(i \in  X_l)
[\prod_{l=1}^k g(X_l) - \prod_{l=1}^k g'(X_l) ]\Big|
\nonumber\\
&&\leq
\sup_{j \in \Lambda_L}
\sum_{k=2}^\infty \frac{1}{k!}  \sum_{X_1, \cdots, X_k \in S_\Lambda}I(j\in  \triangle_k)
\Big|  \sum_{i \in \Lambda_L} h_i s_i^D \prod_{l=1}^kI(i\in  X_l)\Big| \Big|\prod_{l=1}^k g(X_l) - \prod_{l=1}^k g'(X_l) \Big|
 \nonumber\\
&&\leq
\sup_{j \in \Lambda_L}
\sum_{k=2}^\infty \frac{1}{(k-1)!}  \sum_{X_1, \cdots, X_k \in S_\Lambda}I(j \in  X_1)
\Big|  \sum_{i \in \Lambda_L} h_i s_i^D\prod_{l=1}^kI(i \in X_l)\Big|
\Big|\prod_{l=1}^k g(X_l) - \prod_{l=1}^k g'(X_l) \Big|
,  \nonumber
\end{eqnarray}
where  
$I(j \in \triangle_k)  \leq  \sum_{ l=1}^k  I ( j \in X_l) $ and permutation symmetry in the summation over $X_1, \cdots, X_k$ 
have been used.  
The inequality
\begin{equation}
\Big| \prod_{l=1}^k g(X_l) - \prod_{l=1}^k g'(X_l) \Big| \leq \sum_{l=1} ^k \prod_{j=1}^{l-1} |g(X_j)||g(X_l)-g'(X_l) |\hspace{-2mm} \prod_{j=l+1} ^k\hspace{-2mm} |g'(X_j)|,
\label{tri}
\end{equation}
enables us to evaluate the norm as follows: 
\begin{eqnarray}
&&\| F(g)-F(g')\| 
 \\
&&\leq 
\sup_{m \in \Lambda_L}\sum_{k=2}^\infty \frac{1}{(k-1)!}
 \sum_{X_1, \cdots, X_k\in S_\Lambda }I(m \in  X_1)\Big|  \sum_{i \in \Lambda_L} h_i s_i^D\prod_{l=1}^kI(i \in X_l)\Big|
\left[\sum_{l=1} ^k \prod_{j=1}^{l-1} |g(X_j)|
|g(X_l)-g'(X_l) |
 \prod_{j=l+1} ^k
|g'(X_j)|\right]
 \nonumber
\\
&&\leq \sup_{m \in \Lambda_L}
\sum_{k=2}^\infty \frac{1}{(k-1)!}
 \sum_{X_1, \cdots, X_k\in S_\Lambda }I(m \in X_1)\Big|  \sum_{i \in X_1} h_i s_i^D\prod_{l=2}^kI(i\in X_l)\Big|
 \left[\sum_{l=1} ^k \prod_{j=1}^{l-1} |g(X_j)|
|g(X_l)-g'(X_l) |
\prod_{j=l+1} ^k
|g'(X_j)|\right]
\nonumber\\
&&\leq \sup_{m \in \Lambda_L}
\sum_{k=2}^\infty \frac{1}{(k-1)!}
\sup_{i_2, \cdots, i_k\in \Lambda_L}  \sum_{X_1, \cdots, X_k\in S_\Lambda }I(m \in X_1)\Big|  \sum_{i \in X_1} h_i s_i^D\Big|\prod_{l=2}^kI(i_l\in X_l)\nonumber\\
&&\quad\times
\sum_{l=1} ^k \prod_{j=1}^{l-1} |g(X_j)|
|g(X_l)-g'(X_l)|
\prod_{j=l+1} ^k
|g'(X_j)|
\nonumber\\
&&
\leq \sup_{m \in\Lambda_L}
\sum_{k=2}^\infty \frac{1}{(k-1)!}
\Big[  \sum_{X_1\in S_\Lambda}I(m \in  X_1)\Big|  \sum_{i \in X_1} h_i s_i^D\Big||g(X_1)-g'(X_1)|
\prod_{j=2} ^k\sup_{i_j\in \Lambda_L}\sum_{X_j\in S_\Lambda}|g'(X_j)|I(i_j \in X_j)
\nonumber\\
&&+ \sum_{l=2} ^k
 \sum_{X_1\in S_\Lambda}I(m \in  X_1)\Big|  \sum_{i \in X_1} h_i s_i^D\Big| |g(X_1)|
\prod_{j=2}^{l-1}\sup_{i_j\in \Lambda_L}\sum_{X_j\in S_\Lambda }|g(X_j)|I(i_j \in  X_j)
\nonumber\\
&&\quad\times
\sup_{i_l\in \Lambda_L}\sum_{X_l\in S_\Lambda}|g(X_l)-g'(X_l)|I(i_l \in  X_l)
\prod_{j=l+1} ^k\sup_{i_j\in \Lambda_L}\sum_{X_j\in S_\Lambda}|g'(X_j)|I(i_j \in X_j) \Big]
\nonumber\\
&&\leq 
 \sum_{k=2}^\infty \frac{1}{(k-1)!}\Big[ \| g-g'\| \Big( \prod_{j=2} ^k\| g'\|/\Delta\Big)+
 \hspace{-1mm}
 \sum_{l=2} ^k \|g\| \Big( \prod_{j=2}^{l-1}\| g \| /\Delta\Big)  \|g-g'\| /\Delta \Big( \prod_{j=l+1} ^k\| g'\|/\Delta\Big) \Big]
 \nonumber\\&&
= \|g-g'\|\sum_{k=2}^\infty \frac{1}{(k-1)!} \sum_{l=1} ^k (\| g \|/\Delta)^{l-1} (\|g'\|/\Delta)^{k-l} 
 \leq \|g-g'\| \sum_{k=2} ^\infty \frac{k (\delta/\Delta)^{k-1}}{(k-1)!} = [ e^{\delta/\Delta} (1+\delta/\Delta ) -1] \| g-g'\|,
\nonumber
\end{eqnarray}
where  the  energy gap $\Delta>0$ in the unperturbed
model  is defined by
\begin{eqnarray}
\Delta
:=\inf_{X \in S_\Lambda}\Big| \sum_{i \in X} 
 h_is_i^D  \Big|,
\end{eqnarray}
and the following inequality for any $g$
$$
\sup_{j \in \Lambda_L} \sum_{X \in S_\Lambda }I(j \in  X)|g(X)| 
\leq  \frac{\| g\|}{\Delta} \leq  \frac{\delta}{\Delta},
$$
has been used.  
The condition  $\frac{1}{2}=e^{\delta/\Delta} (1+\delta/\Delta ) -1$  fixes $\alpha= \delta/\Delta$.  
To obtain the bound on $\| F(g) \|$,  let us evaluate $\| F(0)\|$ first. Since
$$
F(0) (X)= \frac{-\sum_{b \in B_\Lambda}J_b \delta _{b, X }}{ \sum_{i \in X} h_is_i^D },
$$
the norm is given by
\begin{eqnarray}
\| F(0)\| &=&  \sup_{j \in \Lambda_L}\sum_{X \in S_\Lambda } I(j \in  X)| \sum_{b \in B_\Lambda}
 J_b \delta _{b, X}  | 
 \nonumber \\
 &=&\sup_{j \in \Lambda_L}  \sum_{b \in B_\Lambda}
 |J_b | I(j \in  b) 
=
\sup_{j\in \Lambda_L} \sum_{b \in \partial \{  j \}}
 |J_b|.   
\end{eqnarray}
For $ \sup_{j\in \Lambda_L} \sum_{b \in \partial \{  j \}}
| J_b| \leq \frac{\delta}{2} := \frac{\alpha \Delta}{2} $ and $\| g\| \leq \delta$, 
$$
\| F(g) \| = \| F(g) -F(0) +F(0)\| \leq \| F(g) -F(0)\|  +\| F(0)\| \leq \frac{\| g \|}{2} +  \frac{\delta}{2}\leq \delta.
$$
This completes the proof of Lemma \ref{L2}. $\Box$

\paragraph{Proof of  Theorem \ref{T1}} 
Let  $\alpha > 0$ be the solution of the
 equation $e^{\alpha}(1+\alpha)-1 =\frac{1}{2}$, and define  a positive constant
$\delta := \alpha \inf_{i \in \Lambda_L} 
 |h_i|.$
Lemma \ref{L2} and  the contraction mapping theorem
enable us to prove that the fixed point equation 
$F(g) =g$
has a unique solution $g$, which corresponds to the unperturbed energy eigenstate $|s^D\rangle$
and satisfies  $\| g\| \leq \delta $, if $\sup_{j \in \Lambda_L} \sum_{b \in \partial \{ j \} }| J_b|\leq  \frac{\delta}{2}$.  
$\Box$

\subsection{Expansion for energy gap}
Here, we discuss an energy gap  $E_C-E_D$ between two arbitrary energy eigenstates $|C\rangle$ and $ |D\rangle$.
The eigenvalue equation
$$
H_\Lambda^{\rm xZ}(\sigma, {\bm h},\bm J) |C \rangle =E_C | C\rangle
$$
 is written in 
$$
-( \sum_{b\in B_\Lambda} J_b \sigma_b^z +\sum_{i\in \Lambda_L} h_i \sigma_i^x ) |C \rangle = E_C |C \rangle.
$$
There exists a  real valued function  $\phi : \{1,-1\}^{\Lambda_L} \to {\mathbb  R}$, 
and  the state $|C\rangle$ is represented in 
$$|C\rangle = 2^{-|\Lambda_L|/2}\sum_{\sigma \in \Sigma_\Lambda} \sigma_{D} \psi(\sigma) \phi(\sigma) | \sigma\rangle,$$
where  the function $\psi$ defines the reference state
$$|D \rangle=2^{-|\Lambda_L|/2}\sum_{\sigma \in \Sigma_\Lambda} \sigma_{D} \psi(\sigma) | \sigma \rangle.$$
Assume that in the unperturbed model $\bm J=\bm 0$,  there exists a sub-lattice $C \in S_\Lambda $, such that
$$
|C\rangle = 2^{-|\Lambda_L|/2}\sum_{\sigma \in \Sigma_\Lambda} \sigma_C  | \sigma\rangle.
$$
The eigenvalue equation can be represented in terms of $\phi(\sigma)$.
\begin{equation}
- \sum _{b\in B_\Lambda} J_b\sigma_b \sigma_D\psi(\sigma)  \phi(\sigma) - \sum_{i\in \Lambda_L}h_i \sigma_D^{(i)} \psi(\sigma^{(i)})  \phi(\sigma^{(i)})
 = E_C\sigma_D \psi(\sigma) \phi(\sigma).
\end{equation}
Therefore
\begin{equation}
- \sum _{b\in B_\Lambda} J_b \sigma_b \phi(\sigma)  - \sum_{i\in \Lambda_L} h_i\frac{\sigma_D^{(i)}}{\sigma_D}\frac{\psi(\sigma^{(i)})}{\psi(\sigma)} \phi(\sigma^{(i)})
= E_C \phi(\sigma).
\label{eigeneqpsixZ}
\end{equation}
The following relation
$$
\frac{\sigma^{(i)}_D}{\sigma_D} \frac{\psi(\sigma^{(i)})}{\psi(\sigma)}  =s_i^D \exp \Big[ \sum_{X \in S_\Lambda }  I(i \in X)g(X) \sigma_X  \Big]
$$
and the eigenvalue equation (\ref{eigeneqpsixZ}) give 
\begin{equation}
 \sum _{b \in B_\Lambda} J_b \phi(\sigma) + \sum_{i\in \Lambda_L}h_i s_i^D\phi(\sigma^{(i)})   \exp \Big[\sum_{X \in S_\Lambda }  I(i \in X)g(X) \sigma_X \Big] = -E_C 
 \phi(\sigma).
\end{equation}
The eigenvalue equation with the energy eigenvalue $E_D$ times $\phi(\sigma)$ is 
\begin{equation}
 \sum _{b\in B_\Lambda} J_b \phi(\sigma) + \sum_{i\in \Lambda_L}h_i s_i^D\phi(\sigma)   \exp \Big[\sum_{X \in S_\Lambda }  I(i \in X)g(X) \sigma_X \Big] = -E_D
 \phi(\sigma).
 \end{equation}
The difference between above two equations gives 
\begin{equation} 
- \sum_{i\in \Lambda_L}h_i s_i^D[\phi(\sigma^{(i)}) -\phi(\sigma)]   \exp \Big[\sum_{X \in S_\Lambda }  I(i \in X)g(X) \sigma_X \Big]= (E_C -E_D) \phi(\sigma).
\end{equation}
To obtain the Kirkwood-Thomas equation for the first excited state, represent the function $\phi(\sigma)$ in terms of 
 a real valued  function $f(X)$ of an arbitrary subset $X \in S_\Lambda  $,
\begin{equation}
\phi(\sigma) =\sum_{X \in S_\Lambda } f(X) \sigma_X.
\end{equation}
This gives 
$$
\phi(\sigma^{(i)})  -\phi(\sigma) = -2 \sum_{X \in S_\Lambda } f(X) \sigma_X I(i \in X)
$$
Then we have
\begin{eqnarray}
 2 \sum _{i\in \Lambda_L} h_i s_i^D  \sum_{Y \in S_\Lambda }  I(i \in  Y) f(Y)  \sigma_Y
 \exp \Big[ \sum_{X \in S_\Lambda }  I(i \in  X)g(X) \sigma_X \Big]  =
 (E_C-E_D) \sum_{X\in S_\Lambda } f(X)\sigma_X. \nonumber
\end{eqnarray}
Define  a function $\exp^{(1)}$ by
$$
\exp^{(1)} x := e^x-1 =\sum_{k=1}^\infty\frac{x^k}{k!},
$$
then we have
\begin{eqnarray}
 && \sum_{Y \in S_\Lambda }  \Delta_Y f(Y)  \sigma_Y
+2\sum _{i \in \Lambda_L} h_is_i^D   \sum_{Y \in S_\Lambda }  I(i\in Y) f(Y)  \sigma_Y\exp^{(1)} \Big[ \sum_{X \in S_\Lambda }  I(i \in X)g(X) \sigma_X \Big]   \nonumber \\
 &&=
 (E_C-E_D) \sum_{X\in S_\Lambda } f(X)\sigma_X.
\end{eqnarray}
where an energy gap $\Delta_Y$  for $Y\in \Lambda_L$ is defined by
$$
\Delta _Y:=  2 \sum_{i \in Y} h_is_i^D.
$$
The orthonormalization property (\ref{on})  gives
\begin{eqnarray}
 &&2\sum _{i\in\Lambda_L} h_is_i^D  \sum_{Y \in S_\Lambda }  I(i \in Y) f(Y)\sum_{k=1}^\infty \frac{1}{k!}  \sum_{ X_1, \cdots, X_k \in S_\Lambda}
\delta_{
X_1 \triangle \cdots \triangle X_k, Y\triangle Z} \prod_{l=1}^k g(X_l) I(i \in X_l)  
  \nonumber \\&&=
 (E_C-E_D-\Delta_Z) f(Z).
\end{eqnarray}
Define a function $e(X)$ of a sub-lattice $X \in S_\Lambda $, by 
$e(C):=E_C-E_D-\Delta_C$ for $X=C$ and  $e(X):= f(X)/f(C)$ for $X\neq C$. 

 For $Z=C$, 
 \begin{eqnarray}
 e(C)=2\sum _{i\in \Lambda_L} h_i s_i^D  \sum_{Y \in S_\Lambda }  I(i \in  Y) e(Y)\sum_{k=1}^\infty \frac{1}{k!}  \sum_{ X_1, \cdots, X_k \in S_\Lambda}
\delta_{
X_1 \triangle \cdots \triangle X_k, Y\triangle M} \prod_{l=1}^k g(X_l) I(i \in X_l) =: F(e)(C),
  \nonumber 
\end{eqnarray}
 For $Z\neq C,$
\begin{eqnarray}
 && e(Z)=\frac{1}{\Delta_Z-\Delta_C} \Big[ e(C)  e(Z)\\
 &&-2\sum _{i \in \Lambda_L} h_i s_i^D\sum_{Y \in S_\Lambda }  I(i\in Y) e(Y)\sum_{k=1}^\infty \frac{1}{k!}  \sum_{ X_1, \cdots, X_k \in S_\Lambda}
\delta_{
X_1 \triangle \cdots \triangle X_k, Y\triangle Z} \prod_{l=1}^k g(X_l) I( i \in  X_l)  \Big]=:F(e)(Z).
  \nonumber 
\end{eqnarray}
These two equations define a fixed point equation $F(e)=e$, whose  solution $e$ gives the state $|C\rangle$ except its
normalization. To prove the uniqueness of the solution,
define a norm of the function $e$ by 
\begin{equation}
\| e \| :=|e(C)|+ \sum_{X \in S_\Lambda } |\Delta_X-\Delta_C| |e(X)|.
\label{norm}
\end{equation}
The following theorem implies that  there is no level crossing against 
a sufficiently  small perturbation $\bm J$. \\

{\theorem \label{T1.2}  Consider the transverse field EA model defined by the Hamiltonian 
(\ref{xZ}).
For two different sub-lattices $C, D \in S_\Lambda$, 
let $s^C, s^D \in \{1,-1\}^{\Lambda_L}$ be their corresponding sequences defined by (\ref{sD}). 
If the sequence of exchange interactions $\bm J$ is sufficiently weak,
then  there exists a sufficiently small constant $\delta>0$ depending on the sequence of coupling constants $(\bm J, \bm h)$, such that 
the energy gap $E_C-E_D$ in the perturbed model satisfies
 $$H_\Lambda^{\rm xZ}(s^C, \bm 0, \bm h) - H_\Lambda^{\rm xZ}(s^D, \bm 0, \bm h)-\delta  < E_C -E_D <  H_\Lambda^{\rm xZ}(s^C, \bm 0, \bm h) - H_\Lambda^{\rm xZ}(s^D, \bm 0, \bm h)+\delta,$$
 for almost all $\bm h \in {\mathbb R}^{B_\Lambda}$.}\\

Then, the following lemma can be proven.\\

\noindent 
{\lemma  \label{L3} 
 Consider the model under the conditions in Lemma \ref{L2}, and assume that 
the lower bound on energy gap is sufficiently large. 
There exist a sufficiently small constant $\delta > 0$ and $0 < K < 1$
, such that 
\begin{equation}
\|F(e)-F(e') \| \leq  K \| e-e' \|,   \ \ \| F(e) \| \leq \delta, \ \  {\rm for}\  \| e\|, \| e'\| \leq \delta,
\end{equation} 
for almost all $\bm h \in {\mathbb R}^{\Lambda_L}$.}\\

\noindent
{\bf Proof.}
The difference between  two evaluations of  energy gap 
\begin{eqnarray}
 &&|F(e)(C)-F(e')(C)| \\
 &&=\Big|2\sum _{i\in \Lambda_L} h_is_i^D  \sum_{Y \in S_\Lambda }  I(i\in Y) [e(Y)-e'(Y)]\sum_{k=1}^\infty \frac{1}{k!}  \sum_{ X_1, \cdots, X_k \in S_\Lambda}
\delta_{X_1 \triangle \cdots \triangle X_k, Y\triangle C} \prod_{l=1}^k g(X_l) I(i\in  X_l)\Big| \nonumber \\
&&\leq  \sum_{Y \in S_\Lambda } | \Delta_Y ||e(Y)-e'(Y)|\sum_{k=1}^\infty \frac{1}{k!} 
 \prod_{l=1}^k  \sup_{i_l \in \Lambda_L} \sum_{X_l \in S_\Lambda } |g(X_l) |I(i_l\in  X_l)\nonumber \\
  \nonumber %
  \\&& \leq \sum_{Y \in S_\Lambda }  |\Delta_Y-\Delta_C +\Delta_C| |e(Y)-e'(Y)|\sum_{k=1}^\infty \frac{1}{k!} \Big(\frac{\delta}{\Delta}\Big)^k
  \nonumber \\
  && \leq  \| e-e'\|(1+ |\Delta_C|/\Delta') (e^{\delta/\Delta} -1),
  \end{eqnarray}
  where $\Delta' := \inf_Y |\Delta_Y-\Delta_C|$. 
  \begin{eqnarray}
 && \|F(e)-F(e')\| = |F(e)(C)-F(e')(C)|+
\sum_{Z\in S_\Lambda } \Big| e(C)  e(Z) -e'(C) e'(Z) \nonumber \\
 &&-2\sum _{i \in\Lambda_L} h_is_i^D  \sum_{Y \in S_\Lambda }  I(i\in  Y)[ e(Y)-e'(Y)]\sum_{k=1}^\infty \frac{1}{k!} 
 \sum_{ X_1, \cdots, X_k \in S_\Lambda}
\delta_{
X_1 \triangle \cdots \triangle X_k, Y\triangle Z} \prod_{l=1}^k g(X_l) I(i \in X_l)  \Big|
  \nonumber \\
 &&\leq
  |F(e)(C)-F(e')(C)|+
\sum_{Z\in S_\Lambda } | e(C)  e(Z) -e'(C) e'(Z)| \nonumber \\
 &&+ \sum_{Z\in S_\Lambda }\sum_{Y \in S_\Lambda }\Big| 2 \sum _{i\in Y} h_i s_i^D \Big| | e(Y)-e'(Y)|\sum_{k=1}^\infty \frac{1}{k!}  \sup_{i_1, \cdots, i_k\in \Lambda_L}\sum_{ X_1, \cdots, X_k \in S_\Lambda} 
\delta_{X_1 \triangle \cdots \triangle X_k, Y\triangle Z} \prod_{l=1}^k |g(X_l)| I( i_l \in X_l) 
  \nonumber \\
   &&\leq
 |F(e)(C)-F(e')(C)|+
 | e(C)-e'(C)|\sum_{Z\in S_\Lambda } |e(Z)|  + |e(C)|\sum_{Z\in S_\Lambda } |e(Z)-e'(Z)| \nonumber \\
 &&+   \sum_{Y \in S_\Lambda }\Big| 2\sum _{i\in Y} h_i s_i^D \Big| |e(Y)-e'(Y)|\sum_{k=1}^\infty \frac{1}{k!} \sup_{i_1, \cdots, i_k\in \Lambda_L} \sum_{ X_1, \cdots, X_k \in S_\Lambda}
\prod_{l=1}^k |g(X_l)| I(i_l\in X_l)  
  \nonumber \\
   &&\leq
 |F(e)(C)-F(e')(C)|+
 | e(C)-e'(C)|\delta/\Delta' + \delta\sum_{Z\in S_\Lambda } |e(Z)-e'(Z)| \nonumber \\
 &&+ \sum_{Y \in S_\Lambda }  |\Delta_Y| |e(Y)-e'(Y)|\sum_{k=1}^\infty \frac{1}{k!}  
\prod_{l=1}^k\sup_{i_l \in \Lambda_L} \sum_{  X_l\in S_\Lambda} |g(X_l)| I(i_l\in  X_l)  
  \nonumber \\
   &&\leq \| e-e'\|(1+ |\Delta_C|/\Delta') (e^{\delta/\Delta} -1)+\delta/\Delta'  \|e-e'\| 
 + (1+|\Delta_C|/\Delta')(e^{\delta/\Delta}-1)\| e-e'\|
  \nonumber \\
 && =[2 (1+|\Delta_C|/\Delta')(e^{\delta/\Delta}-1)+\delta/\Delta' ]\| e-e'\| = K \| e-e'\|.
\end{eqnarray}
If $K < 1$, $F$ is a contraction mapping.
  This completes the proof of Lemma \ref{L3}. $\Box$

\paragraph{Proof of  Theorem \ref{T1.2}} 
 For two different sub-lattices $C \neq D$, 
 Lemma \ref{L3} and the contraction mapping theorem imply that  unique function $e: 2^{\Lambda_L}\to{\mathbb R}$  exists and satisfies
  $|E_C-E_D -\Delta_C|=:  |e(C)| \leq \delta$ for sufficiently small $\delta>0$. 
Therefore,  the energy gap $E_C-E_D$ in the perturbed model satisfies
 $$H_\Lambda^{\rm xZ}(s^C, \bm 0, \bm h) - H_\Lambda^{\rm xZ}(s^D, \bm 0, \bm h)-\delta  <E_C-E_D <  H_\Lambda^{\rm xZ}(s^C, \bm 0, \bm h) - H_\Lambda^{\rm xZ}(s^D, \bm 0, \bm h)+\delta,$$
 for almost all $\bm J \in {\mathbb R}^{B_\Lambda}$.
This completes the proof of  Theorem \ref{T1.2}. $\Box$

\section{Transverse field EA model around the EA model}
Define  discrete  transformation 
 \begin{equation}
 P_w:= \sigma_{\Lambda_L}^w= \prod_{i\in \Lambda_L} \sigma_i^w, \label{P}
\end{equation}
for $w=x,y,z$.
The Hamiltonian (\ref{xZ}) is invariant
$$P_xH_\Lambda^{\rm xZ}(\bm \sigma, \bm h, \bm J)P_x=H_\Lambda^{\rm xZ}(\bm \sigma, \bm h, \bm J),$$  
for $w=x$. 
The unitary operator $P_x$ transforms
each spin to  $\sigma _i^z \mapsto P_x \sigma_i ^z P_x =-\sigma^z_i$  and $\sigma _i^x 
\mapsto P_x  \sigma_i ^x P_x=\sigma^x_i$, $\sigma _i^y \mapsto P_x  \sigma_i ^y P_x =-\sigma^y_i$, 
and this symmetry
 corresponds to ${\mathbb Z}_2$  symmetry in the EA model for $\bm h = \bm 0$. 
 Define a state $| \sigma\rangle$ with a sequence of eigenvalues  $\sigma \in  \{1,-1\}^{\Lambda_L}$ of spin operators 
 $(\sigma_i^z)_{i\in \Lambda_L}$  by
 $$
 \sigma_i^z | \sigma \rangle = \sigma_i |\sigma\rangle.
 $$
To remove the trivial  two-fold degeneracy 
due to the global ${\mathbb Z}_2$ symmetry,
assume a symmetry breaking condition at an arbitrarily fixed site $i_0 \in \Lambda_L$,  such that 
\begin{equation}\sigma_{i_0}=1.
\label{BC}
\end{equation}  
Define sub-lattice  $\Lambda_L' := \Lambda_L\setminus \{i_0\}$.
Any  ${\mathbb Z}_2$ symmetry  broken state $|\sigma\rangle$  is given by  $\sigma \in  \{1,-1\}^{ \Lambda_L' }.$

\subsection{Unperturbed system}
The following lemma guarantees  the non-degenerate property of energy eigenstates under the condition (\ref{BC})
in the classical EA model defined by  the Hamiltonian (\ref{Z}). 
In the following, we study eigenstates of the Hamiltonian (\ref{xZ}) under the condition (\ref{BC}) in a convergent perturbative expansion.\\

\noindent
{\lemma \label{L4} Consider  the unperturbed model  defined by the Hamiltonian (\ref{Z}) 
 in $d$-dimensional hyper cubic lattice $\Lambda_L$.
For  any two different eigenvalue sequences  $\sigma, \sigma'\in \{1,-1 \}^ {\Lambda_L'} $ of  the operator 
sequence $(\sigma_i^z)_{i\in \Lambda_L}$
satisfying the fixed spin condition  (\ref{BC}), the Hamiltonian takes different values 
$$
H_\Lambda^{Z}(\sigma,\bm J) \neq H_\Lambda^{Z}(\sigma',\bm J),
$$
for almost all $\bm J \in {\mathbb R}^{B_\Lambda}$}.\\

\noindent
{\bf Proof. }
Let  $b_1, b_2, b_3, \cdots  \in B_\Lambda$ be a sequence of
 bonds,  such that   $i_0  \in b_1$ and  each collection $\{ b_n | n \leq N \}$ is connected for an arbitrary positive integer $N$.   
Define a connected  sub-lattice  $X_N ( \in S_\Lambda ) $ for this bond sequence $(b_n)_{n\leq N}$ by
$$
X_N:=   \bigcup_{n=1}^N b_n .
$$
The following mathematical inductivity with respect to $N$ enables us to prove this lemma.\\
For $N=1$,   $X_1=b_1=\{ i_0,i_1\}$  is a single bond. The condition $\sigma_{i_0}=1$ implies
 $$
 H_{X_1} ^Z(\sigma, \bm J) =- J_{b_1} \sigma_{i_1}.
 $$
 Since the Hamiltonian takes different values 
 for $\sigma_{i_1} =\pm1$,  this lemma is valid for $N=1$.\\ 
 For an arbitrary positive integer $N$, assume the validity of this lemma. Then, 
 \begin{equation}
H_{X_N}^{Z}(\sigma,\bm J) \neq H_{X_N}^{Z}(\sigma',\bm J),
\label{N}
\end{equation}
is valid for any two different configurations $\sigma, \sigma' \in \{1,-1\}^{ X_{N} }$ satisfying (\ref{BC}) for almost all $\bm J$. \\
For $N+1$, 
 let $\sigma, \sigma'\in \{1,-1\}^{ X_{N+1}}$ be two different configurations satisfying (\ref{BC}).
Consider
the  equation  for $J_{b_{N+1}}$
\begin{equation}
H_{X_{N+1}}^{Z}(\sigma,\bm J)  = H_{X_{N+1}}^{Z}(\sigma',\bm J), 
\label{eqJ}
\end{equation}
which has the following representation in terms of $H_{X_N}$ and $J_{b_{N+1}}$
$$
H_{X_N}^Z(\sigma |_{X_N},\bm J)  - J_{b_{N+1}} \sigma_{b_{N+1}} = H_{X_N}^Z(\sigma' |_{X_N},\bm J)  - J_{b_{N+1}} \sigma'_{b_{N+1}}.
$$
Since 
$|X_{N+1}\setminus X_N| \leq 1$,   $\sigma_{b_{N+1}} =\sigma'_{b_{N+1}}$ 
implies 
the assumption (\ref{N}).  Then, 
the equation (\ref{eqJ}) has no solution for $\sigma_{b_{N+1}} =\sigma'_{b_{N+1}}$. For $ \sigma_{b_{N+1}} -\sigma'_{b_{N+1}} = \pm 2$,
 the corresponding solutions of the equation (\ref{eqJ}) are given by
\begin{equation}
J_{b_{N+1}} =\pm  \frac{1}{2}[H_{X_N}^Z(\sigma' |_{X_N},\bm J) - H_{X_N}^Z(\sigma |_{X_N},\bm J) ].
\label{solJ}
\end{equation}
Therefore, 
 $$
H_{X_{N+1}}^{Z}(\sigma,\bm J)  \neq  H_{X_{N+1}}^{Z}(\sigma',\bm J), 
$$ 
is valid also for  $N+1$  for almost all $J_{b_{N+1}} \in {\mathbb R}$ except  the solutions (\ref{solJ}).  Then, this lemma is valid for an arbitrary positive integer $N$.
This completes the proof. $\Box$

\subsection{Expansion method}
To obtain an arbitrary energy eigenstate in the transverse EA model, consider the following unitary transformed Hamiltonian 
\begin{equation}
\tilde H_\Lambda(\sigma, {\bm h},\bm J) :=U H_\Lambda^{\rm xZ} U^\dag= -\sum_{b\in B_\Lambda} J_b \sigma_b^x -\sum_{i\in \Lambda_L}h_i\sigma_i^z,
\end{equation}
where $U\sigma_i^xU^\dag=\sigma_i^z$ and $U\sigma_i^zU^\dag=-\sigma_i^x$.
The discrete unitary transformation defined by (\ref{P}) is transformed into 
$$ P_z = \sigma_{\Lambda_L}^z = U^\dag P_x U $$
 In the unperturbed model for $\bm h= \bm 0$, Lemma \ref{L4}  guarantees non-degeneracy of energy eigenstates satisfying  
 the condition (\ref{BC}). 
 For an arbitrary  sub-lattice $D \ ( \in S_\Lambda )$, the corresponding eigenstate in the unperturbed model is represented in
 $$
  2^{-|\Lambda_L|/2} \sum_{\sigma \in \Sigma_\Lambda}\sigma_D| \sigma \rangle,
 $$
 and another degenerate eigenstate  is
 $$
2^{-|\Lambda_L|/2} \sum_{\sigma \in \Sigma_\Lambda}\sigma_{D^c}| \sigma \rangle.
$$
Note that the unitary transformed  energy eigenstate in the unperturbed model can be represented in
$$
U |s^D \rangle=\sum_{\sigma \in \{1,-1\}^{\Lambda_L}}\sigma_D| \sigma \rangle,
$$where  a sequence $(s_i^D)_{i\in \Lambda_L}$  is defined by (\ref{sD}).
For the perturbed model, let $\psi_\pm(\sigma)$ be a function $\psi_\pm : \{-1,1\}^{\Lambda_L} \to {\mathbb R}$ of spin configuration,
and express the energy eigenstate of the Hamiltonian with respect to the unperturbed state
$$
|\pm \rangle=2^{-(|\Lambda_L|+1)/2} \sum_{\sigma \in \Sigma_\Lambda}\sigma_D (1\pm \sigma_{\Lambda_L}) \psi_\pm(\sigma)| \sigma \rangle.
$$
These are eigenstates of $ P_z$ satisfying $ P_z |\pm \rangle = \pm|\pm\rangle.$
Note that $\psi_\pm(\sigma) =1$ for $\bm h= \bm 0$. 
The normalization condition   $\langle \pm |\pm\rangle=1$ 
gives 
\begin{equation}
\sum_{\sigma \in \Sigma_\Lambda} \psi_\pm(\sigma)^2 = 2^{|\Lambda_L|}.
\label{normalization2}
\end{equation}
The eigenvalue equation defined by
$$
\tilde H_\Lambda(\sigma, {\bm h},\bm J) |\pm \rangle =E_\pm |\pm \rangle
$$
 is written in 
$$
-( \sum_{b\in B_\Lambda} J_b \sigma_b^x +\sum_{i\in \Lambda_L} h_i \sigma_i^z ) |\pm \rangle = E_\pm |\pm\rangle.
$$
If $\sigma_i ^x |\sigma \rangle = | \tau \rangle$,  $\tau_i = - \sigma_i $ and $\tau _j=\sigma_j$ for $j\neq i$. 
$$
\sigma_b^x| \sigma\rangle =\sigma_i^x \sigma_j^x| \sigma\rangle
= |\sigma^{(i,j)} \rangle,
 $$
where $\sigma^{(i,j)}$  denotes a  spin configuration replaced  by $(\sigma_i , \sigma_j) \to (-\sigma_i, -\sigma_j)$. 
This eigenvalue equation can be represented in terms of $\psi(\sigma)$.
\begin{equation}
 \sum _{b\in B_\Lambda} J_b\sigma_D^{(b)} (1\pm\sigma_{\Lambda_L}^{(b)})  \psi_\pm(\sigma^{(b)})  + \sum_{i\in \Lambda_L}h_i \sigma_i \sigma_D(1\pm\sigma_{\Lambda_L})  \psi_\pm(\sigma) 
 = -E_\pm\sigma_D(1\pm\sigma_{\Lambda_L})  \psi_\pm(\sigma).
\end{equation}
Therefore
\begin{equation}
(1\pm\sigma_{\Lambda_L}) \Big[ \sum _{b\in B_\Lambda} J_b  \frac{\sigma_D^{(b)}\psi_\pm(\sigma^{(b)})}{\sigma_D\psi_\pm(\sigma)}  + \sum_{i\in \Lambda_L} h_i\sigma_i 
+E_\pm\Big]=0.
\label{eigeneqpsiZx}
\end{equation}
This equation determines 
 $\psi_\pm(\sigma)$ as a function of  $\sigma$ satisfying $\sigma_{\Lambda_L} =\pm 1.$  
To obtain the Kirkwood-Thomas equation for the state, represent the functions $\psi_\pm(\sigma)$ in terms of 
 a real valued  function $g(X)$ of an arbitrary non-empty sub-lattice $X \in S_\Lambda $,
\begin{equation}
\psi_\pm(\sigma) =\frac{1}{2}(1\pm \sigma_{\Lambda_L}) \exp \Big[-\frac{1}{2} \sum_{X \in S_\Lambda } g(X) \sigma_X \Big].
\end{equation}
Note the following relations 
\begin{equation}
\psi_\pm(\sigma^{(b)}) =\frac{1}{2}(1\pm \sigma_{\Lambda_L}) \ \exp \Big[ -\frac{1}{2}\sum_{X \in S_\Lambda } g(X) \sigma_X +\sum_{X \in S_\Lambda }  I(b \in \partial X)g(X) \sigma_X  \Big].
\end{equation}
$$
 \frac{\psi_\pm(\sigma^{(b)})}{\psi_\pm(\sigma)}  =\frac{1}{2}(1\pm \sigma_{\Lambda_L}) \  \exp \Big[ \sum_{X \in S_\Lambda }  I(b \in \partial X)g(X) \sigma_X  \Big].
$$
Define 
$$
s_b^D :=\frac{\sigma^{(b)}_D}{\sigma_D}. 
$$
These and the eigenvalue equation (\ref{eigeneqpsiZx}) give 
\begin{equation}
\frac{1}{2}(1\pm \sigma_{\Lambda_L}) \Big( \sum _{b\in B_\Lambda} J_bs_b^D   \exp \Big[\sum_{X \in S_\Lambda }  I(b \in \partial X)g(X) \sigma_X \Big] + \sum_{i\in \Lambda_L}h_i \sigma_i +E_\pm \Big) =0.
\end{equation}
We expand the exponential function in power series. The first order  term in the exponential function  is  given by
\begin{equation}
 \sum _{b\in B_\Lambda} J_b s_b^D \sum_{X \in S_\Lambda }  I(b \in \partial X)g(X) \sigma_X=
 \sum_{X \in S_\Lambda }  \sum _{b \in \partial X} J_bs_b^D g(X)  \sigma_X ,
\end{equation}
then we have
\begin{eqnarray}
 &&\frac{1}{2}(1\pm \sigma_{\Lambda_L}) \Big(\sum _{b\in B_\Lambda} J_bs_b^D +E_\pm +  \sum_{X \in S_\Lambda }  \sum _{b \in \partial X} J_b s_b^D g(X)  \sigma_X\nonumber \\
 &&+\sum _{b\in B_\Lambda} J_b s_b^D \exp^{(2)} \Big[ \sum_{X \in S_\Lambda }  I(b \in \partial X)g(X) \sigma_X \Big] + \sum_{i\in \Lambda_L}h_i \sigma_i \Big)=0,
\end{eqnarray}
where 
$$
\exp^{(2)} x := e^x-1-x =\sum_{k=2}^\infty\frac{x^k}{k!}.
$$
The orthonormalization property (\ref{on}) 
  gives
\begin{eqnarray}
E_\pm =-\sum _{b\in B_\Lambda} J_bs_b^D  -\sum _{c\in B_\Lambda} J_c s_c^D\sum_{k=2}^\infty \frac{1}{k!}  \sum_{ X_1, \cdots, X_k\in S_\Lambda}
\delta_{
X_1 \triangle \cdots \triangle X_k, \phi} \prod_{l=1}^k g(X_l)I(c\in \partial X_l) \nonumber \\
 \mp\sum _{c\in B_\Lambda} J_c s_c^D\sum_{k=2}^\infty \frac{1}{k!}  \sum_{ X_1, \cdots, X_k\in S_\Lambda}
\delta_{
X_1 \triangle \cdots \triangle X_k, \Lambda_L} \prod_{l=1}^k g(X_l)I(c\in \partial X_l),
\label{Epm}
\end{eqnarray}
and  $g$ should satisfy
\begin{eqnarray}
g(X)&=&\frac{-1}{\sum _{b\in \partial X} J_bs_b^D} \left[\sum _{c\in B_\Lambda} J_c s_c^D\sum_{k=2}^\infty \frac{1}{k!}  \sum_{ X_1, \cdots, X_k \in S_\Lambda}
\delta_{
X_1 \triangle \cdots \triangle X_k, X} \prod_{l=1}^k g(X_l) I(c\in \partial X_l)\right.\nonumber\\
&&\left.+\sum_{i \in \Lambda_L} h_i \delta _{X, \{i\}} \right]
=: F(g)(X),
\end{eqnarray}
for $X \neq \phi, \Lambda_L$.  The normalization (\ref{normalization2}) fixes $g(\phi)\pm g(\Lambda_L)$.
The first term in the energy eigenvalue 
 is identical to the energy of the  spin configuration $s^D$ for $h_i=0$.
To obtain the  energy eigenstate in the transverse field EA model with a given $\bm J$
for a sufficiently small  $h:= \sup_{c\in B_\Lambda}\sum_{i \in c} |h_i|$,
define a norm of the function $g(X)$  with  a constant $M>0$  by 
\begin{equation}
\| g \| := \sup_{c\in B_\Lambda}\sum_{X \in S_\Lambda } I(c\in \partial X) \Big| \sum_{b\in \partial X}J_b s_b^D\Big|   |g(X)|
(h M)^{-w(X)},
\label{norm}
\end{equation}
where $w(X)$ is defined by the cardinality of the smallest connected sub-lattice that contains $X (\in S_\Lambda )$. 
We say that a sub-lattice  $X$ is connected, if for any $i,j \in X$ there exists a sequence $i_1, i_2, \cdots, i_n \in X$,  such that 
$i_1=i$, $i_n =j$ and $\{i_k, i_{k+1}\} \in B_\Lambda$ for $k=1,   \cdots, n-1$.
Then, the following theorem can be proven.\\

{\theorem  \label{T2} Consider the transverse field EA model defined by the Hamiltonian
(\ref{xZ}) with sufficiently weak coupling constants $\bm  h$.
Let $D (\in S_\Lambda )$ be an arbitrary sub-lattice.
There are two energy eigenstates with energy eigenvalues $E_\pm$
 corresponding to the two-fold degenerate energy eigenstates $|s^D\rangle \pm |s^{D^c}\rangle$ 
 in the unperturbed model, such that the energy gap $|E_+-E_-|$ is exponentially small in the system size $|\Lambda_L|$
 for almost all $\bm J \in {\mathbb R}^{B_\Lambda}$. 
 }\\
 
Theorem \ref{T2} is proven by the following lemma and the contraction mapping theorem.

\noindent
{\lemma  \label{L5}
There exist a constant  $\delta>0$ and define $M:=\frac{2}{\delta}$
, such that if $
 hM <  1$,
\begin{equation}
\|F(g)-F(g') \| \leq   \| g-g' \|/2,   \ \ \| F(g) \| \leq \delta, \ \ {\rm for}\  \| g\|, \| g'\| \leq \delta,
\end{equation} 
 for almost all $\bm J \in {\mathbb R}^{B_\Lambda}$. 
Then, there exists a constant  $A >0$ depending on $\delta$, such that 
  the energy gap $|E_+-E_-|$ satisfies the following exponentially small  bound
   $$ |E_+-E_-| \leq  A \Big|\sum_{b\in B_\Lambda}J_b s_b^D \Big| (h M)^{|\Lambda_L|}.$$
 }\\

\noindent
{\bf Proof.} 
  The norm $\| F(g)-F(g') \|$ is represented in
\begin{eqnarray}
&&\| F(g)-F(g')\| 
=\sup_{c \in B_\Lambda}\sum_{X\in S_\Lambda }I(c\in \partial X)
\Big| \sum_{b\in B_\Lambda} J_b s_b^D 
\sum_{k=2}^\infty \frac{1}{k!}  \sum_{X_1, \cdots, X_k \in S_\Lambda}
\delta_{\triangle_k, X}\nonumber\\
&&\quad\times[ \prod_{l=1}^k g(X_l) - \prod_{l=1}^k g'(X_l) ]
\prod_{l=1}^kI(b\in \partial X_l)
(hM)^{-w(X)}
\Big| \nonumber\\
&&\leq
\sup_{c \in B_\Lambda}
\sum_{k=2}^\infty \frac{1}{k!}  \sum_{X_1, \cdots, X_k \in S_\Lambda}I(c\in \partial \triangle_k)\Big|\sum_{b\in B_\Lambda} J_b s_b^D \prod_{l=1}^kI(b\in \partial X_l)\Big|
\Big|\prod_{l=1}^k g(X_l) - \prod_{l=1}^k g'(X_l) \Big|
(hM)^{-w(\triangle_k)}
 \nonumber\\
&&\leq
\sup_{c \in B_\Lambda}
\sum_{k=2}^\infty \frac{1}{(k-1)!}  \sum_{X_1, \cdots, X_k \in S_\Lambda}I(c\in \partial X_1)\Big|\sum_{b\in B_\Lambda} J_b s_b^D
\prod_{l=1}^kI(b\in \partial X_l) \Big|
\Big|\prod_{l=1}^k g(X_l) - \prod_{l=1}^k g'(X_l) \Big|
(hM)^{-w(\triangle_k)}
,  \nonumber
\end{eqnarray}
where
$I(c\in \partial \triangle_k)  \leq  \sum_{ l=1}^k  I ( c \in  \partial X_l) $
and permutation symmetry in the summation over $X_1, \cdots, X_k$ 
have been used. 
An inequality (\ref{tri}) and
$w(\triangle_k) \leq \sum_{l=1}^k w(X_l) $
enable us to evaluate the norm as follows: 
\begin{eqnarray}
\| F(g)-F(g')\| 
&&\leq 
\sup_{c \in B_\Lambda}\sum_{k=2}^\infty \frac{1}{(k-1)!}
 \sum_{X_1, \cdots, X_k\in S_\Lambda }I(c\in \partial X_1)\Big|\sum_{b\in B_\Lambda} J_b s_b^D \prod_{l=1}^kI(b\in \partial X_l)\Big|\nonumber\\
&&\quad\times\left[\sum_{l=1} ^k \prod_{j=1}^{l-1} |g(X_j)|
|g(X_l)-g'(X_l) |
 \prod_{j=l+1} ^k
|g'(X_j)|\right](hM)^{-w(\triangle_k)} 
\nonumber\\
&&= \sup_{c \in B_\Lambda}
\sum_{k=2}^\infty \frac{1}{(k-1)!}
 \sum_{X_1, \cdots, X_k\in S_\Lambda }I(c\in \partial X_1)\Big|\sum_{b\in \partial X_1} J_b s_b^D\prod_{l=2}^kI(b\in \partial X_l) \Big|\nonumber\\
&&\quad\times\left[
\sum_{l=1} ^k \prod_{j=1}^{l-1} |g(X_j)|
|g(X_l)-g'(X_l) |
\prod_{j=l+1} ^k
|g'(X_j)|\right](hM)^{-\sum_{j=1}^kw(X_j)}
\nonumber\\
&&\leq
\sum_{k=2}^\infty \frac{1}{(k-1)!}
\sup_{b_1, \cdots, b_k \in \Lambda_L} \sum_{X_1, \cdots, X_k\in S_\Lambda }\Big|\sum_{b \in \partial X_1} J_b s_b^D \Big|\nonumber\\
&&\quad\times\left[
\sum_{l=1} ^k \prod_{j=1}^{l-1} |g(X_j)|I(b_j\in \partial X_j)(h M)^{-w(X_j)}
|g(X_l)-g'(X_l)|I(b_l\in \partial X_l)(hM)^{-w(X_l)}
\right.\nonumber\\
&&\left.\qquad\times
 \hspace{-2mm} 
\prod_{j=l+1} ^k
 \hspace{-2mm} 
|g'(X_j)|I(b_j\in \partial X_j)(h M)^{-w(X_j)}
\right].
\end{eqnarray}
The norm is bounded in terms of norms of $g, g'$
\begin{eqnarray}
&&\| F(g)-F(g')\|  \nonumber \\
&&\leq 
\sum_{k=2}^\infty \frac{1}{(k-1)!} \Big[ \|g-g'\| \prod_{j=2} ^k\| g'\|/\Delta+\hspace{-1mm}
 \sum_{l=2} ^k\|g\| \prod_{j=2}^{l-1}\| g \| /\Delta \|g-g'\| /\Delta \prod_{j=l+1} ^k\| g'\|/\Delta\Big]\nonumber\\
&&= \|g-g'\|\sum_{k=2}^\infty \frac{1}{(k-1)!} \sum_{l=1} ^k (\| g \|/\Delta)^{l-1} (\|g'\|/\Delta)^{k-l} \nonumber \\
&&\leq \|g-g'\| \sum_{k=2} ^\infty \frac{k (\delta/\Delta)^{k-1}}{(k-1)!} =[ e^{\delta/\Delta} (1+\delta/\Delta ) -1] \| g-g'\|,
\nonumber
\end{eqnarray}
where  the  energy gap $\Delta>0$ from the state in the unperturbed
model  is defined by
\begin{eqnarray}
\Delta
:=\inf_{
X \in S_\Lambda }  \Big| \sum_{b\in \partial X}J_b s_b^D \Big|,
\end{eqnarray}
and  the following inequality for $\| g \|  \leq \delta$,
$$
\sup_{c\in B_\Lambda} \sum_{X \in S_\Lambda }I(c\in \partial X)|g(X)| (hM)^{-{w(X)}} 
\leq  \frac{\| g\|}{\Delta}\leq  \frac{\delta}{\Delta},
$$
has been used.
The condition  $\frac{1}{2}=K=e^{\delta/\Delta} (1+\delta/\Delta ) -1$
fixes $\alpha  = \delta /\Delta$. 
To obtain the bound on $\| F(g) \|$,  let us evaluate $\| F(0)\|$ first. Since
$$
F(0) (X)= \frac{-\sum_{i\in \Lambda_L}h_i\delta _{X, \{i\}} }{ \sum_{b \in \partial X} J_b s_b^D },
$$
the norm is given by
\begin{eqnarray}
\| F(0)\| &=&  \sup_{c\in B_\Lambda}\sum_{X \in S_\Lambda } I(c\in \partial X)| \sum_{i\in \Lambda_L}
 h_i \delta _{X, \{i\}}  | (h M)^{-w(X)} 
 \nonumber \\
 &=&  \sup_{c\in B_\Lambda} \sum_{i\in \Lambda_L}\sum_{X \in S_\Lambda } I(c\in \partial X) 
 |h_i |\delta _{X, \{i\}}  (h M)^{-w(X)}
\nonumber  \\
&=& \sup_{c\in B_\Lambda}\sum_{i \in \Lambda_L} I(c\in \partial \{i \} )
| h_i| (h M)^{-1}	
 =\sup_{c\in B_\Lambda}\sum_{i \in c} | h_i|(hM)^{-1} =M^{-1}.
\end{eqnarray}
Define  $
M:= \frac{ 2}{\delta} := \frac{2}{\alpha  \Delta}$,  and then
$$
\| F(g) \| = \| F(g) -F(0) +F(0)\| \leq \| F(g) -F(0)\|  +\| F(0)\| \leq \frac{\| g \|}{2} +  \frac{\delta}{2}\leq \delta.
$$
The energy gap 
\begin{equation}
E_+-E_-=2\sum _{c\in B_\Lambda} J_c s_c^D\sum_{k=2}^\infty \frac{1}{k!}  \sum_{ X_1, \cdots, X_k\in S_\Lambda}
\delta_{X_1 \triangle \cdots \triangle X_k, \Lambda_L} \prod_{l=1}^k g(X_l)I(c\in \partial X_l),
\end{equation}
can be evaluated in terms of the norm $\| g\|$ using an inequality $\sum_{l=1}^kw(X_k) \geq  |\Lambda_L|$ for $\triangle_k = \Lambda_L $
\begin{eqnarray}
|E_+-E_-|&=&\Big| 2\sum _{c\in B_\Lambda} J_c s_c^D\sum_{k=2}^\infty \frac{1}{k!}  \sum_{ X_1, \cdots, X_k\in S_\Lambda}
\delta_{X_1 \triangle \cdots \triangle X_k, \Lambda_L} \prod_{l=1}^k g(X_l)I(c\in \partial X_l)\Big| \nonumber \\
&=&\Big| 2\sum _{c\in B_\Lambda} J_c s_c^D\sum_{k=2}^\infty \frac{1}{k!}  \sum_{ X_1, \cdots, X_k\in S_\Lambda}
\delta_{X_1 \triangle \cdots \triangle X_k, \Lambda_L} (hM)^{\sum_{l=1}^kw(X_l)}
\prod_{l=1}^k g(X_l)(hM)^{-w(X_l)}I(c\in \partial X_l)\Big| \nonumber \\
&\leq&\Big| 2\sum _{c\in B_\Lambda} J_c s_c^D\Big| (hM)^{w(\Lambda_L)}\sum_{k=2}^\infty \frac{1}{k!}  \prod_{l=1}^k 
\sup_{c_l\in B_\Lambda}  \sum_{  X_l \in S_\Lambda}| g(X_l)|I(c_l\in \partial X_l)(hM)^{-w(X_l)}\nonumber \\
&\leq&\Big| 2\sum _{c\in B_\Lambda} J_c s_c^D\Big| (hM)^{|\Lambda_L|}\sum_{k=2}^\infty \frac{1}{k!}  \Big(\frac{\| g\|}{\Delta}\Big)^k\nonumber \\
&\leq&\Big| 2\sum _{c\in B_\Lambda} J_c s_c^D\Big| (hM)^{|\Lambda_L|}\exp^{(2)}\frac{\delta}{\Delta}=
A \Big|  \sum _{c\in B_\Lambda} J_c s_c^D\Big| (hM)^{|\Lambda_L|}
,
\end{eqnarray}
where  $A$ is defined by
$A:=\exp^{(2)} \alpha.$ 
This completes the proof of Lemma \label{L5}. $\Box$

\paragraph{Proof of  Theorem \ref{T2}.}  Let  $\alpha > 0$ be the solution of an equation $e^{\alpha}(1+\alpha)-1 =\frac{1}{2}$, and 
define a constant  $\delta >0$ by
\begin{equation}
\delta := \alpha \inf_{
X \in S_\Lambda } \Big|  \sum_{b\in \partial X}J_b s_b^D \Big|.
\end{equation}
Lemma \ref{L5} and  the contraction mapping theorem
enable us to prove that the fixed point equation 
$F(g) =g$
has unique solution $g$ satisfying $\| g\| \leq \delta $, if $h:=\sup_{c \in B_\Lambda}\sum_{i\in c} |h_i| < \frac{\delta}{2}$.
This solution gives the energy eigenstates 
  \begin{equation}
 | \pm  \rangle =2^{-(|\Lambda_L|+1)/2}  \sum_{\sigma\in \{1,-1\}^{\Lambda_L}} \sigma_D (1\pm\sigma_{\Lambda_L})\exp \Big[-\frac{1}{2}\sum_{X\in S_\Lambda} g(X)  \sigma_X \Big] |\sigma\rangle, 
 \label{ketpm1}
 \end{equation}
 corresponding to $|D\rangle\pm|D^c\rangle$ in the unperturbed model.
The energy gap $|E_+-E_-|$ satisfies the following exponentially small  bound
 $$ |E_+-E_-| \leq  \exp^{(2)} \alpha \Big|\sum_{b\in B_\Lambda}J_b s_b^D \Big| (h M)^{|\Lambda_L|}=
    \exp^{(2)} \alpha \Big|\sum_{b\in B_\Lambda}J_b s_b^D \Big| (2h/\delta)^{|\Lambda_L|}
.$$
for $M:=\frac{2}{\delta}.$
This completes the proof of Theorem \ref{T2} $\Box$

\subsection{Expansion for energy gap}
Here, we discuss  an energy eigenstate $|\pm \rangle'$ different from the sate $|\pm\rangle$ 
obtained in Lemma \ref{L5}.
The eigenvalue equation
$$
\tilde H_\Lambda(\sigma, {\bm h},\bm J) |\pm \rangle' =E_\pm' | \pm \rangle'
$$
is written in 
$$
-( \sum_{b\in B_\Lambda} J_b \sigma_b^x +\sum_{i\in \Lambda_L} h_i \sigma_i^z ) |\pm \rangle' = E_\pm' |\pm \rangle'.
$$
There exists a  real valued function  $\phi_\pm : \{1,-1\}^{\Lambda_L} \to {\mathbb  R}$, 
and  the state $|\pm \rangle'$ is represented in 
$$|\pm\rangle' =2^{-|\Lambda_L|/2} \sum_{\sigma \in \Sigma_\Lambda} \sigma_{D}(1\pm\sigma_{\Lambda_L}) \psi_\pm(\sigma) \phi_\pm(\sigma) | \sigma\rangle,$$
where  the function $\psi$ defines the reference state
$$|\pm \rangle=2^{-(|\Lambda_L|+1)/2}\sum_{\sigma \in \Sigma_\Lambda} \sigma_{D}(1\pm\sigma_{\Lambda_L}) \psi_\pm(\sigma)  | \sigma \rangle.$$
Assume that  for the unperturbed model $\bm h=\bm 0,$  there exists a sub-lattice $C \in S_\Lambda $, such that
$$
 |\pm \rangle' = 2^{-(|\Lambda_L|+1)/2} \sum_{\sigma \in \Sigma_\Lambda} \sigma_{C} (1\pm\sigma_{\Lambda_L})| \sigma\rangle.
$$
The eigenvalue equation can be represented in terms of $\phi(\sigma)$.
\begin{equation}
(1\pm\sigma_{\Lambda_L}) \Big[ \sum _{b\in B_\Lambda} J_b\sigma_D^{(b)}  \psi_\pm(\sigma^{(b)})  \phi_\pm (\sigma^{(b)}) + \sum_{i\in \Lambda_L}h_i \sigma_i \sigma_D \psi_\pm(\sigma)  \phi_\pm(\sigma)
 + E_\pm \sigma_D\psi_\pm(\sigma) \phi_\pm(\sigma) \Big]=0.
\label{eigeneqpsi}
\end{equation}
Therefore
\begin{equation}
(1\pm\sigma_{\Lambda_L})  \Big[ \sum _{b\in B_\Lambda} J_b  \frac{\sigma_D^{(b)}\psi_\pm(\sigma^{(b)})}{\sigma_D\psi_\pm(\sigma)} \phi_\pm(\sigma^{(b)}) + \sum_{i\in \Lambda_L} h_i\sigma_i  \phi_\pm(\sigma)
+ E_\pm \phi_\pm(\sigma) \Big]=0.
\label{eigeneqpsi2}
\end{equation}
The following relation
$$
\frac{\sigma^{(b)}_D}{\sigma_D} \frac{\psi_\pm(\sigma^{(b)})}{\psi_\pm(\sigma)}  =s_b^D  \exp \Big[ \sum_{X \in S_\Lambda }  I(b \in \partial X)g(X) \sigma_X  \Big]
$$
and the eigenvalue equation (\ref{eigeneqpsi2}) give 
\begin{equation}
(1\pm\sigma_{\Lambda_L}) \Big[ \sum _{b\in B_\Lambda} J_bs_b^D \phi_\pm(\sigma^{(b)})   \exp \Big[\sum_{X \in S_\Lambda }  I(b \in \partial X)g(X) \sigma_X \Big] + \sum_{i\in \Lambda_L}h_i \sigma_i\phi_\pm(\sigma) +E_\pm' 
 \phi_\pm(\sigma)\Big]=0.
\end{equation}
The energy eigenvalue equation with $E_\pm$ times $\phi(\sigma)$ is 
\begin{equation}
(1\pm\sigma_{\Lambda_L}) \Big[ \sum _{b\in B_\Lambda} J_bs_b^D \phi_\pm(\sigma)   \exp \Big[\sum_{X \in S_\Lambda }  I(b \in \partial X)g(X) \sigma_X \Big] + \sum_{i\in \Lambda_L}h_i \sigma_i\phi_\pm(\sigma) +E_\pm 
 \phi_\pm(\sigma)\Big]=0.
\end{equation}
The difference between above two equations gives 
\begin{equation}
(1\pm\sigma_{\Lambda_L}) \Big[ \sum _{b\in B_\Lambda} J_bs_b^D[ \phi_\pm(\sigma^{(b)})  -\phi_\pm(\sigma)] \exp \Big[\sum_{X \in S_\Lambda }  I(b \in \partial X)g(X) \sigma_X \Big]  + (E_\pm' -E_\pm) \phi_\pm(\sigma)\Big]=0.
\end{equation}
To obtain the Kirkwood-Thomas equation for the first excited state, represent the function $\phi(\sigma)$ in terms of 
 a real valued  function $f(X)$ of an arbitrary subset $X \in S_\Lambda  $,
\begin{equation}
\phi_\pm(\sigma) =\frac{1}{2}(1\pm\sigma_{\Lambda_L}) \sum_{X \in S_\Lambda } f(X) \sigma_X.
\end{equation}
This gives 
$$
\phi_\pm(\sigma^{(b)})  -\phi_\pm(\sigma) = -(1\pm\sigma_{\Lambda_L})  \sum_{X \in S_\Lambda } f(X) \sigma_X I(b \in \partial X)
$$
Then we have
\begin{eqnarray}
&&\Big[ 2 \sum _{b\in B_\Lambda} J_bs_b^D  \sum_{Y \in S_\Lambda }  I(b \in \partial Y) f(Y)  \sigma_Y
 \exp \Big[ \sum_{X \in S_\Lambda }  I(b \in \partial X)g(X) \sigma_X \Big] \nonumber \\
&&  -(E_\pm'-E_\pm) \sum_{X\in S_\Lambda } f(X)\sigma_X \Big] (1\pm\sigma_{\Lambda_L}) =0. 
\end{eqnarray}
Define  a function $\exp^{(1)}$ by
$$
\exp^{(1)} x := e^x-1 =\sum_{k=1}^\infty\frac{x^k}{k!},
$$
then we have
\begin{eqnarray}
 &&\Big[ \sum_{Y \in S_\Lambda }  \Delta_Y f(Y)  \sigma_Y
+2\sum _{b\in B_\Lambda} J_b s_b^D  \sum_{Y \in S_\Lambda }  I(b\in \partial Y) f(Y)  \sigma_Y\exp^{(1)} \Big[ \sum_{X \in S_\Lambda }  I(b \in \partial X)g(X) \sigma_X \Big]   \nonumber \\
 &&-
 (E_\pm'-E_\pm) \sum_{X\in S_\Lambda } f(X)\sigma_X \Big](1\pm\sigma_{\Lambda_L}) =0.
\end{eqnarray}
where an energy gap $\Delta_Y$  for $Y\in \Lambda_L$ is defined by
$$
\Delta _Y:=  2 \sum_{b \in \partial Y} J_bs_b^D.
$$
The orthonormalization property (\ref{on})  gives
\begin{eqnarray}
 &&2\sum _{b\in B_\Lambda} J_b s_b^D  \sum_{Y \in S_\Lambda }  I(b\in \partial Y) f(Y)\sum_{k=1}^\infty \frac{1}{k!}  \sum_{ X_1, \cdots, X_k \in S_\Lambda}
(\delta_{
X_1 \triangle \cdots \triangle X_k\triangle  Y\triangle Z,\phi}
 \nonumber \\&&
  \pm \delta_{X_1 \triangle \cdots \triangle X_k\triangle  Y\triangle Z,\Lambda_L}) \prod_{l=1}^k g(X_l) I(b\in \partial X_l)  
- (E_\pm '-E_\pm -\Delta_{Z})[ f(Z)\pm f(Z^c) ] =0.
\end{eqnarray}
Therefore, 
\begin{eqnarray}
 &&2\sum _{b\in B_\Lambda} J_b s_b^D  \sum_{Y \in S_\Lambda }  I(b\in \partial Y)[ f(Y)\pm f(Y^c)]\sum_{k=1}^\infty \frac{1}{k!}  \sum_{ X_1, \cdots, X_k \in S_\Lambda}  \delta_{X_1 \triangle \cdots \triangle X_k\triangle  Y\triangle Z,\phi}
 \nonumber \\&&
 \times \prod_{l=1}^k g(X_l) I(b\in \partial X_l)  
- (E_\pm '-E_\pm -\Delta_{Z})[ f(Z)\pm f(Z^c) ] =0.
\end{eqnarray}
Define a function $e_\pm(X)$ of non-empty sub-lattice $X \ (\in S_\Lambda )$ by
\begin{equation}
e_\pm(X) :=
\left\{
\begin{array}{lll}
E_\pm ' -E_\pm - \Delta_C                & ( \ X=C \ ) \\
\frac{ f(X) \pm f(X^c)}{f(C) \pm f(C^c)} & ( \  X\neq C \ ) 
\end{array}
\right.
\label{e}
\end{equation}
For $Z=C$, 
 \begin{eqnarray}
 e_\pm(C) = \sum_{Y \in S_\Lambda } 2\sum _{b\in \partial Y} J_b s_b^D  e_\pm(Y)\sum_{k=1}^\infty \frac{1}{k!}  \sum_{ X_1, \cdots, X_k \in S_\Lambda}
\delta_{
X_1 \triangle \cdots \triangle X_k\triangle Y, C} \prod_{l=1}^k g(X_l) I(b\in \partial X_l) =: F( e_\pm)(C),
  \nonumber 
\end{eqnarray} 
For $Z\neq C,$
\begin{eqnarray}
 && e_\pm(Z)=\frac{1}{\Delta_Z-\Delta_C} \Big[ e_\pm(C)  e_\pm(Z)\\
 &&- \sum_{Y \in S_\Lambda }  2\sum _{b\in \partial Y} J_b s_b^D e_\pm(Y)\sum_{k=1}^\infty \frac{1}{k!} 
  \sum_{ X_1, \cdots, X_k \in S_\Lambda}
\delta_{
X_1 \triangle \cdots \triangle X_k\triangle Y, Z} \prod_{l=1}^k g(X_l) I(b\in \partial X_l)  \Big]=:F(e_\pm)(Z).
  \nonumber 
\end{eqnarray} 
These equations define a fixed point equation $F(e)=e $
, whose  solution $e$ gives the state $|\pm\rangle'$ except its normalization
. To prove the uniqueness of the solution,
define a norm for the function $e_\pm$ by 
\begin{equation}
\|e_\pm \| :=|e_\pm(C) |+ \sum_{X \in S_\Lambda } |\Delta_X-\Delta_C| |e_\pm(X)|.
\label{norm}
\end{equation}
The following theorem implies that  there is no level crossing against 
a sufficiently  small perturbation $\bm h$. \\

{\theorem \label{T2.2} Consider the random transverse field EA model defined by the Hamiltonian (\ref{xZ}) .
For two different sub-lattices $C, D \in S_\Lambda$, 
let $s^C, s^D \in \{1,-1\}^{\Lambda_L}$ be their corresponding sequences defined by (\ref{sD}). 
If the sequence of transverse fields $\bm h$ is sufficiently weak, there exists a sufficiently small constant $\delta>0$ depending on the sequence of coupling constants $(\bm J, \bm \epsilon)$, such that 
the energy gap $E_\pm'-E_\pm$ in the perturbed model satisfies
 $$H_\Lambda^{\rm xZ}(s^C, \bm J, \bm 0) - H_\Lambda^{\rm xZ}(s^D, \bm J, \bm 0)-\delta  < E_\pm' -E_\pm <  H_\Lambda^{\rm xZ}(s^C, \bm J, \bm 0) - H_\Lambda^{\rm xZ}(s^D, \bm J, \bm 0)+\delta,$$
 for almost all $\bm J \in {\mathbb R}^{B_\Lambda}$.}\\
 
Theorem \ref{T2.2} is proven by the following lemma.\\

\noindent
{\lemma  \label{L6} Consider the model under the conditions in Lemma \ref{L5}, 
and assume  that  lower bound on  energy gap 
$
\inf_{X,Y \in S_\Lambda }
 |\Delta_X-\Delta_Y| 
$  
in the unperturbed model
 is sufficiently large.
There exist  constants $\delta > 0$ and $0 < K < 1$
, such that 
\begin{equation}
\|F(e_\pm)-F(e_\pm') \| \leq  K \| e_\pm-e_\pm' \|,   \ \ \| F(e_\pm) \| \leq \delta, \ \  {\rm for}\  \| e_\pm\|, \| e_\pm'\| \leq \delta,
\end{equation} 
for almost all $\bm J \in {\mathbb R}^{B_\Lambda}$.}\\

\noindent
{\bf Proof.}     For lighter notation, we remove indices $\pm$ from $e_\pm$.  The difference between  two evaluations of  energy gap 
\begin{eqnarray}
 |F(e)(C)-F(e')(C)| 
 &&
 =\Big|2\sum _{b\in B_\Lambda} J_b s_b^D  \sum_{Y \in S_\Lambda }  I(b\in \partial Y) [e(Y)-e'(Y)]
 \nonumber \\
&&\times\sum_{k=1}^\infty \frac{1}{k!}  \sum_{ X_1, \cdots, X_k \in S_\Lambda}
\delta_{
X_1 \triangle \cdots \triangle X_k\triangle Y, C} \prod_{l=1}^k g(X_l) I(b\in \partial X_l)\Big| \nonumber \\
&&\leq  \sum_{Y \in S_\Lambda }  |\Delta_Y| |e(Y)-e'(Y)|
 \sum_{k=1}^\infty \frac{1}{k!} 
 \prod_{l=1}^k  \sup_{b_l \in B_\Lambda} \sum_{X_l \in S_\Lambda } |g(X_l) |I(b_l\in \partial X_l)\nonumber \\
  \nonumber %
  \\&& \leq \sum_{Y \in S_\Lambda }  |\Delta_Y-\Delta_C +\Delta_C| |e(Y)-e'(Y)|\sum_{k=1}^\infty \frac{1}{k!} \Big(\frac{\delta}{\Delta}\Big)^k
  \nonumber \\
  && \leq  \| e-e'\|(1+ |\Delta_C|/\Delta') (e^{\delta/\Delta} -1),
  \end{eqnarray}
  where $\Delta' := \inf_Y |\Delta_Y-\Delta_C|$.  
  \begin{eqnarray}
 && \|F(e)-F(e')\| = |F(e)(C)-F(e')(C)|+
\sum_{Z\in S_\Lambda } \Big| e(C)  e(Z) -e'(C)e'(Z) \nonumber \\
 &&-2\sum _{b\in B_\Lambda} J_b s_b^D  \sum_{Y \in S_\Lambda }  I(b\in \partial Y)[ e(Y)-e'(Y)]\sum_{k=1}^\infty \frac{1}{k!}  \sum_{ X_1, \cdots, X_k \in S_\Lambda}
\delta_{
X_1 \triangle \cdots \triangle X_k\triangle Y,  Z} \prod_{l=1}^k g(X_l) I(b\in \partial X_l)  \Big|
  \nonumber \\
 &&\leq
  |F(e)(C)-F(e')(C)|+
\sum_{Z\in S_\Lambda } | e(C)  e(Z) -e'(C) e'(Z)| \nonumber \\
 &&+   \sum_{Y \in S_\Lambda }\sum_{Z\in S_\Lambda } \Big| 2\sum _{b\in  \partial Y} J_b s_b^D \Big| | e(Y)-e'(Y)|\sum_{k=1}^\infty \frac{1}{k!} \sup_{b_1, \cdots, b_k \in B_\Lambda} \sum_{ X_1, \cdots, X_k \in S_\Lambda}
\delta_{X_1 \triangle \cdots \triangle X_k\triangle Y, Z} \prod_{l=1}^k |g(X_l)| I(b_l\in \partial X_l) 
  \nonumber \\
   &&\leq
 |F(e)(C)-F(e')(C)|+
 | e(C)-e'(C)|\sum_{Z\in S_\Lambda } |e(Z)|  + |e(C)|\sum_{Z\in S_\Lambda } |e(Z)-e'(Z)| \nonumber \\
 &&+   \sum_{Y \in S_\Lambda } \Big| 2\sum _{b\in \partial Y } J_b s_b^D \Big| |e(Y)-e'(Y)|\sum_{k=1}^\infty \frac{1}{k!}  \sup_{b_1, \cdots, b_k \in B_\Lambda} \sum_{ X_1, \cdots, X_k \in S_\Lambda}
\prod_{l=1}^k |g(X_l)| I(b_l\in \partial X_l)  
  \nonumber \\
   &&\leq
 |F(e)(C)-F(e')(C)|+
 |e(C)-e'(C)|\delta/\Delta' + \delta\sum_{Z\in S_\Lambda } |e(Z)-e'(Z)| \nonumber \\
 &&+ \sum_{Y \in S_\Lambda } | \Delta_Y| |e(Y)-e'(Y)|\sum_{k=1}^\infty \frac{1}{k!}  
\prod_{l=1}^k\sup_{c_l \in B_\Lambda} \sum_{  X_l\in S_\Lambda} |g(X_l)| I(c_l\in \partial X_l)  
  \nonumber \\
   &&\leq \| e-e'\|(1+ |\Delta_C|/\Delta') (e^{\delta/\Delta} -1)+\delta/\Delta'  \|e-e'\| 
 + (1+|\Delta_C|/\Delta')(e^{\delta/\Delta}-1)\| e-e'\|
= K \| e-e'\| , 
\end{eqnarray}
where $K:=2 (1+|\Delta_C|/\Delta')(e^{\delta/\Delta}-1)+\delta/\Delta' .$
If $K < 1$, $F$ is a contraction mapping.
  This completes the proof of Lemma \ref{L6}. $\Box$

\paragraph{Proof of Theorem \ref{T2.2}}
Lemma \ref{L6} implies that $$|E_\pm'-E_\pm- \Delta_C| = |e(C)|\leq  \delta. $$
This implies  
$$H_\Lambda^{\rm xZ}(s^C, \bm J, \bm 0) - H_\Lambda^{\rm xZ}(s^D, \bm J, \bm 0)-\delta  < E_\pm' -E_\pm <  H_\Lambda^{\rm xZ}(s^C, \bm J, \bm 0) - H_\Lambda^{\rm xZ}(s^D, \bm J, \bm 0)+\delta,$$
 for almost all $\bm J \in {\mathbb R}^{B_\Lambda}$.
This completes the proof of  Theorem \ref{T2.2}. $\Box$

\subsection{Remarks on results for  the random transverse field EA model}  
\paragraph{1.
}  For the random transverse field EA model,   
the Perron-Frobenius theorem  enables us to prove the uniqueness of the ground state. 
For an arbitrary sequence $(h_i)_{i\in \Lambda_L}$, define
 a sequence $\bm \theta := (\theta_i)_{i\in \Lambda_L}$  
 by $\theta_i =0$ for $h_i > 0$ and $\theta_i=\frac{\pi}{2}$ for $h_i < 0$.
The following  unitary transformation
$$U_{\bm \theta}:= \exp \Big(  i \sum_{i \in \Lambda_L} \theta_i \sigma_i^z\Big)
$$  
transforms the perturbation Hamiltonian 
$$U_{\bm \theta} \sum_{i \in \Lambda_L} h_i \sigma_i ^x U_{\bm \theta}^\dag =  \sum_{i \in \Lambda_L} |h_i| \sigma_i ^x.$$
For the Hamiltonian defined by (\ref{xZ}), the transformed 
 Hamiltonian $H_{\bm \theta}:= U_{\bm \theta} H_\Lambda^{\rm xZ} U_{\bm \theta}^\dag$
 satisfies  
(i) non-positivity of all off-diagonal matrix elements $\langle \sigma |H_{\bm \theta} | \tau  \rangle \leq 0$ for $\sigma \neq \tau$
 and
(ii) connectivity condition $\langle \sigma | H_{\bm \theta}^n |\tau \rangle \neq 0$ for any $\sigma \neq \tau$ for some positive integer $n$.
These conditions allow the application of the Perron-Frobenius theorem to the transverse field EA model.  
This theorem implies that the unique ground state is given by $|GS\rangle = \sum_{\sigma \in \Sigma_\Lambda} \psi(\sigma) |\sigma \rangle$  with positive coefficients
$\psi(\sigma)$.
This result is consistent to Theorem \ref{T1} and \ref{T2} in the present paper.  
This Perron-Frobenius argument  is valid also for the EA model under 
an arbitrary sequence of vector-valued  fields $(\vec h_i)_{i\in \Lambda_L} :=(h_i^x,h_i^y,h_i^z)_{i\in \Lambda_L}.$

As far as the ground state in the random transverse field EA model
is concerned, our expansion method cannot yield new results other than that obtained in 
the Perron-Frobenius argument. For excited states,  however, our method  concludes that there is no
level crossing between any energy eigenstates
against  sufficiently weak perturbations, if we impose a condition to break ${\mathbb Z}_2$ symmetry. 
Our expansion method is applicable also to general perturbation Hamiltonian, which does not satisfy the Perron-Frobenius conditions.  
For example, the uniqueness of the ground state can be shown in 
the EA Hamiltonian  perturbed by sufficiently  
weak transverse exchange interactions  in the following.
\paragraph{2.
} Consider a problem to obtain the ground sate in the EA model by the quantum annealing with the transverse field EA model. Since Theorem  \ref{T2.2} cannot guarantee
 the absence of level crossing between energy eigenvalues $E_+$ and $E_-$,  
 a condition to break $\mathbb Z_2$ in the EA model is necessary to obtain 
 precise solution of the ground sate. 
Theorem \ref{T2.2}  guarantees  that the EA model has no level crossing for sufficiently 
weak transverse field, if there is no $\mathbb Z_2$ symmetry.
\paragraph{3.
}Consider the transverse field EA model with weak transverse fields.   Theorem \ref{T2}
claims that  an  arbitrary  energy gap $|E_+-E_-|$ between split energy eigenvalues  is exponentially small
 in the system size $|\Lambda_L|$.   This fact suggests a spontaneous symmetry  breaking of the $\mathbb Z_2$ 
symmetry in this model.  In our expansion method, however, any results cannot be concluded in the infinite-volume limit

\section{The random bond Heisenberg XYZ model}

\subsection{Energy eigenstate}
Let us consider the  random bond Heisenberg XYZ model defined by the Hamiltonian (\ref{XYZ}), which is invariant 
$$P_wH_\Lambda^{\rm XYZ}(\bm \sigma, \bm h, \bm J)P_w^\dag=H_\Lambda^{\rm XYZ}(\bm \sigma, \bm h, \bm J),$$  
for any  $w=x, y, z$. $P_x$ corresponds to the ${\mathbb Z}_2$ invariance in the EA model.
To construct a perturbative expansion method,  we define several notations.
For an arbitrary sub-lattice $D \in S_{\Lambda}$,  define an eigenvalue sequence $s^D\in \{1,-1 \}^{ \Lambda_L } $ by (\ref{sD})
In the unperturbed model defined by the Hamiltonian $H_\Lambda^Z(\bm \sigma, \bm J) $,  there is a one-to-one correspondence  
between a sub-lattice $D \in S_\Lambda$ and  an arbitrary energy eigenstate $|s^D\rangle$, which breaks ${\mathbb Z}_2$ symmetry. 
To obtain all energy eigenstates in the random bond Heisenberg XYZ model, consider the following unitary transformed Hamiltonian 
\begin{equation}
\tilde H_\Lambda(\bm \sigma,\bm J, {\bm \epsilon}) :=U H_\Lambda^{\rm XYZ} U^\dag= -\sum_{b\in B_\Lambda}( J_b \sigma_b^x +\epsilon_b^x\sigma_b^z+\epsilon_b^y\sigma_b^y),
\label{tildehamil}
\end{equation}
where $U\sigma_i^xU^\dag=\sigma_i^z$ and $U\sigma_i^zU^\dag=-\sigma_i^x$.
Let $\psi_\pm(\sigma)$ be  functions $\psi_\pm: \{-1,1\}^{\Lambda_L} \to {\mathbb R}$ of spin eigenvalues,
and express an eigenstate of the Hamiltonian $\tilde H_\Lambda $ corresponding to $s^D$
$$
|\pm \rangle=2^{-(|\Lambda_L| + 1)/2} \sum_{\sigma \in \{1,-1\}^{\Lambda_L}}\sigma_D(1\pm \sigma_{\Lambda_L}) \psi_\pm(\sigma)| \sigma \rangle.
$$
Note the eigenvalue equation $P_z|\pm\rangle = \pm|\pm\rangle.$
The normalization  of  the sate $|\pm\rangle$ requires 
\begin{equation}
 \sum_{\sigma \in \{1,-1\}^{\Lambda_L}}\psi_\pm(\sigma) ^2=2^{|\Lambda_L|}.
 \label{normalization3}
\end{equation}
Note that 
$\psi_\pm(\sigma) =1$ for $\bm \epsilon= \bm 0$  is given by  a spin configuration $s^D \in \{1,-1\}^{\Lambda_L}$ 
regarded as an energy eigenstate in the EA model for $\bm \epsilon =\bm 0$.
The eigenvalue equation defined by
$$
\tilde H_\Lambda(\bm \sigma,\bm J, {\bm \epsilon}) |\pm \rangle =E_\pm |\pm \rangle
$$
 is written in 
$$
-\sum_{b\in B_\Lambda} (J_b \sigma_b^x +\epsilon_b^x\sigma_b^z- \epsilon_b^y\sigma_b^z \sigma_b^x)  |\pm \rangle 
= E_\pm |\pm \rangle,
$$
where $\sigma_j^y =- i \sigma_j^z \sigma_j^x $ has been used.
If $\sigma_i ^x |\sigma \rangle = | \tau \rangle$,  $\tau_i = - \sigma_i $ and $\tau _j=\sigma_j$ for $j\neq i$. 
$$
\sigma_b^x| \sigma\rangle =\sigma_i^x \sigma_j^x| \sigma\rangle
= |\sigma^{(i,j)} \rangle,
 $$
where $\sigma^{(i,j)}$  denotes a  spin configuration replaced  by $(\sigma_i , \sigma_j) \to (-\sigma_i, -\sigma_j)$. 
This eigenvalue equation can be represented in terms of $\psi(\sigma)$.
\begin{equation}
(1\pm\sigma_{\Lambda_L})\Big[ \sum _{b\in B_\Lambda} [J_b\sigma_D^{(b)}  \psi_\pm(\sigma^{(b)})  + \epsilon^x_b \sigma_b \sigma_D \psi_\pm(\sigma) -
 \epsilon_b^y \sigma_b \sigma_D^{(b)}  \psi_\pm(\sigma^{(b)}) ]+ E_\pm\sigma_D \psi_\pm(\sigma) \Big]=0.
\label{eigenpsi0}
\end{equation}
Therefore
\begin{equation}
(1\pm\sigma_{\Lambda_L})\Big[ \sum _{b\in B_\Lambda}[ (J_b - \epsilon_b^y \sigma_b) \frac{\sigma_D^{(b)}\psi_\pm(\sigma^{(b)})}{\sigma_D\psi_\pm(\sigma)}  +\epsilon_b^x \sigma_b]
+E_\pm \Big]=0.
\label{eigenpsi}
\end{equation}
To obtain the Kirkwood-Thomas equation for  energy eigenstates, represent the function $\psi(\sigma)$ in terms of 
 a real valued  function $g(X)$ of an arbitrary sub-lattice $X \in S_\Lambda $,
\begin{equation}
\psi_\pm(\sigma) =\frac{1}{2}(1\pm\sigma_{\Lambda_L}) \exp \Big[-\frac{1}{2} \sum_{X \in S_\Lambda} g(X) \sigma_X \Big].
\end{equation}
Note the following relations 
\begin{equation}
\psi_\pm (\sigma^{(b)}) =\frac{1}{2}(1\pm\sigma_{\Lambda_L}) \exp \Big[ -\frac{1}{2}\sum_{X \in S_\Lambda} g(X) \sigma_X +\sum_{X \in S_\Lambda}  I(b \in \partial X)g(X) \sigma_X  \Big].
\end{equation}
$$
 \frac{\psi_\pm(\sigma^{(b)})}{\psi_\pm(\sigma)}  = \frac{1}{2}(1\pm\sigma_{\Lambda_L}) \exp \Big[ \sum_{X \in S_\Lambda}  I(b \in \partial X)g(X) \sigma_X  \Big].
$$
Note also, 
$$
\frac{\sigma^{(b)}_D}{\sigma_D}=s^D_b. 
$$
These and the eigenvalue equation (\ref{eigenpsi}) give 
\begin{equation}
\frac{1}{2}(1\pm\sigma_{\Lambda_L})\Big( \sum _{b\in B_\Lambda}\Big[ (J_b- \epsilon_b^y \sigma_b)s^D_b   \exp \Big[\sum_{X \in S_\Lambda}  I(b \in \partial X)g(X) \sigma_X \Big] + \epsilon^x_b \sigma_b \Big]+E_\pm \Big) =0.
\end{equation}
We expand the exponential function in power series. The first order  term in the exponential function gives
\begin{equation}
 \sum _{b\in B_\Lambda}J_b  s^D_b \sum_{X \in S_\Lambda}  I(b \in \partial X)g(X) \sigma_X=
 \sum_{X \in S_\Lambda}  \sum _{b \in \partial X} J_b s^D_b g(X)  \sigma_X ,
\end{equation}
then we have
\begin{eqnarray}
 &&\frac{1}{2}(1\pm\sigma_{\Lambda_L})\Big(\sum _{b\in B_\Lambda}[J_b  s^D_b+(\epsilon_b^x - \epsilon_b^y  s^D_b)\sigma_b] +E_\pm +  \sum_{X \in S_\Lambda} \sum _{b \in \partial X} J_b s^D_b g(X)  \sigma_X
 \label{eigeneqg} \\ &&+\sum _{b\in B_\Lambda}J_b s^D_b \exp^{(2)} \Big[  \sum_{X \in S_\Lambda} I(b \in \partial X)g(X) \sigma_X \Big]
   -\sum _{b\in B_\Lambda} \epsilon_b^y \sigma_b  s^D_b \exp^{(1)} \Big[  \sum_{X \in S_\Lambda} I(b \in \partial X)g(X) \sigma_X \Big] \Big)=0, \nonumber 
\end{eqnarray}
where for a positive integer $n$, the function is defined by 
$$
\exp^{(n)} x := \sum_{k=n}^\infty\frac{x^k}{k!}.
$$
The equation (\ref{eigeneqg}) summed over all $\sigma\in \{1,-1\}^{\Lambda_L'}$ and the orthonormalization property (\ref{on})  give
\begin{eqnarray}
E_\pm =- \frac{1}{2}\sum _{b\in B_\Lambda} J_bs^D_b  &-&\frac{1}{2}\sum_{k=2}^\infty \frac{1}{k!}  \sum_{ X_1, \cdots, X_k\in S_\Lambda }\sum _{c\in B_\Lambda} J_c \delta_{\triangle_k, \phi} s_c^D \prod_{l=1}^k g(X_l)I(c\in \partial X_l) \nonumber \\
 &+&\frac{1}{2}\sum_{k=1}^\infty \frac{1}{k!}  \sum_{ X_1, \cdots, X_k\in S_\Lambda }
\sum _{c\in B_\Lambda} \epsilon_c^y \delta_{c\triangle  \triangle_k,\phi}s_c^D \prod_{l=1}^k g(X_l)I(c\in \partial X_l), \nonumber \\
&\mp& \frac{1}{2}\sum_{k=2}^\infty \frac{1}{k!}  \sum_{ X_1, \cdots, X_k\in S_\Lambda }\sum _{c\in B_\Lambda} J_c 
\delta_{\triangle_k, \Lambda_L} s_c^D \prod_{l=1}^k g(X_l)I(c\in \partial X_l) \nonumber \\
 &\pm& \frac{1}{2}\sum_{k=1}^\infty \frac{1}{k!}  \sum_{ X_1, \cdots, X_k\in S_\Lambda }
\sum _{c\in B_\Lambda} \epsilon_c^y \delta_{c\triangle  \triangle_k,\Lambda_L}s_c^D \prod_{l=1}^k g(X_l)I(c\in \partial X_l).
\label{ED}
\end{eqnarray}
where 
The first term in the right hand side of (\ref{ED})
 for the energy eigenvalue $E_\pm$
is identical to that of the spin configuration $s^D$ for $\bm \epsilon= \bm 0$.
For  $X \neq \phi,  \Lambda_L$, the summation of the equation (\ref{eigeneqg}) multiplied by $\sigma_Y$
over $\sigma\in \{1,-1\}^{\Lambda_L}$ and  the orthonormalization property (\ref{on})  yield
\begin{eqnarray}
g(X)&=&\frac{-1}{\sum _{b\in \partial X} J_bs^D_b} \left[\sum _{c\in B_\Lambda}(\epsilon_c^x - \epsilon_c^y  s_c^D) 
\delta _{c, X}+\sum_{k=2}^\infty \frac{1}{k!}  \sum_{ X_1, \cdots, X_k \in S_\Lambda}
\sum _{c\in B_\Lambda} J_c \delta_{\triangle_k, X}  s_c^D\prod_{l=1}^k g(X_l) I(c\in \partial X_l)\right.\nonumber\\
&&\left.-\sum_{k=1}^\infty \frac{1}{k!}  \sum_{ X_1, \cdots, X_k \in S_\Lambda}
\sum _{c\in B_\Lambda}  \epsilon_c^y \delta_{c \triangle X, \triangle_k}s_c^D\prod_{l=1}^k g(X_l) I(c\in \partial X_l)
 \right]
=: F(g)(X),
\end{eqnarray}
which is a fixed point equation for $g$.  The normalization (\ref{normalization3}) fixes $g(\phi) \pm g(\Lambda_L)$.
 For $\bm \epsilon = \bm 0$,  $2 \sum_{b \in \partial X} J_b s^D_b $ represents
the energy gap of a spin configuration given by  $X:=\{ i \in \Lambda_L'| \sigma_i \neq s^D_i \}$.
Lemma \ref{L4} guarantees  $ \sum_{b \in \partial X} J_b s^D_b \neq 0$ for $X\neq D$
.
\\
 We provide the following  theorem  for the energy eigenstates in 
 the random bond Heisenberg XYZ model on the basis of  convergent perturbative expansions around $\bm \epsilon= \bm 0$.   \\

To prove the convergence of expansion for the energy eigenstate corresponding to the unperturbed state $|s^D\rangle$
 in the random bond XYZ model with a given $\bm J$
for sufficiently small $\bm \epsilon$, define a norm for the function $g(X)$ with positive constants 
$\epsilon:= \sup_{c\in B_\Lambda} \sum_{b\in \partial c} |\epsilon_b^x -\epsilon_b^y s_b^D| $ and  $M>0$  by 
\begin{equation}
\| g \| := \sup_{c\in B_\Lambda}\sum_{X \in S_\Lambda} I(c\in \partial X)\Big| \sum_{b\in \partial X}J_b s^D_b \Big| 
 |g(X)|(\epsilon M)^{-w(X)},
\label{normXYZ}
\end{equation}
where $w(X)$ is the cardinality of the smallest set of connected bonds whose union contains $X$. 
We say that two bonds $\{ i_1, i_2 \}, \{j_1, j_2 \}  \in B_\Lambda$ are connected, if 
$\inf_{1\leq k,l \leq 2}|i_k-j_l| \leq 1$.

\noindent
{\theorem  \label{T3}  Consider  the  Heisenberg XYZ model defined by the Hamiltonian (\ref{XYZ}).
 For  an arbitrary sub-lattice  $D \in S_\Lambda$,
 the  energy eigenstate $ | \pm  \rangle$ with energy eigenvalues $E_\pm$ corresponding
 to the unperturbed energy eigenstate $|s^D\rangle \pm |s^{D^c}\rangle$ exist and
satisfy $\sigma_{\Lambda_L}^z | \pm\rangle = \pm| \pm \rangle$ for almost all $\bm J\in {\mathbb R}^{B_\Lambda}$, 
 if the XY-exchange coupling constants $\bm \epsilon$ are sufficiently weak.  
 The energy gap $|E_+-E_-|$  is exponentially small  in the system size $|\Lambda_L|$
 .}\\

The following lemma and the contraction mapping theorem enable us to prove Theorem \ref{T3}.\\

\noindent
{\lemma \label{L6B} Consider  the  Heisenberg XYZ model defined by the Hamiltonian (\ref{XYZ}).  For an arbitrary sub-lattice $D \in S_\Lambda$, let $s^D\in \{1,-1\}^{\Lambda_L}$
 be its corresponding sequence defined by (\ref{sD}).
 Define $\Delta, \epsilon_y, \epsilon > 0$ by 
  \begin{equation}\Delta:= \inf_{X\in S_\Lambda}\Big|2 \sum_{b\in \partial X}J_b s_b^D \Big|,
\label{K2} 
 \ \ \  \epsilon_y := \Delta\sup_{c\in B_\Lambda, X \in S_\Lambda}  I (c\in \partial X)\sum_{b\in \partial X} | \epsilon_b^y| \Big/ \Big| \sum_{b\in \partial X}  J_b s^D_b\Big|, \ \ \ \epsilon := \sup_{c\in B_\Lambda} \sum_{b\in \partial c} |\epsilon_b^y s_b^D-\epsilon_b^x| .
 \end{equation}
For a constant $\delta>0$,  define   $M := 2/\delta$. 
 If $\epsilon$ and $ \delta $ 
 satisfy    
\begin{equation} e^{2\delta/\Delta}[2\delta/\Delta+1 +(\epsilon M)^{-1 } \epsilon_y /\Delta (2\delta/\Delta+4d-1 ) ] \leq \frac{3}{2},
\ \ \ \  \epsilon < \delta /2, \label{L6C}
\end{equation}
then  the following norms are bounded by     
\begin{equation}
\|F(g)-F(g') \| \leq   \| g-g' \|/2,   \ \  \| F(g) \| \leq \delta, \ \ {\rm for}\  \| g\|, \| g'\| \leq \delta,
\end{equation} 
for almost all $\bm J \in {\mathbb R}^{B_\Lambda}$. 
This unique  function $g$ defines 
the  energy eigenstates $ | \pm  \rangle$ with energy eigenvalues $E_\pm$ corresponding
 to the unperturbed energy eigenstate $|s^D\rangle \pm |s^{D^c}\rangle$
 for almost all $\bm J$. 
There exists  a constant $A>0$ depending on $\delta > 0$, such that the energy gap $|E_+-E_-|$  has an exponentially small bound
 $$
 |E_+-E_-| \leq  A\Big(\Big| \sum _{c\in B_\Lambda} J_c  s_c^D \Big| + \sum _{c\in B_\Lambda} |\epsilon_c^y | \Big)  (\epsilon M)^{|\Lambda_L|},
 $$
}

\noindent
{\bf Proof.} %
  The norm $\| F(g)-F(g') \|$ is represented in
\begin{eqnarray}
&&\| F(g)-F(g')\| 
= \sup_{c \in B_\Lambda} \sum_{X\in S_\Lambda}I(c\in \partial X)\Big|
\sum_{k=2}^\infty \frac{1}{k!}  \sum_{X_1, \cdots, X_k \in S_\Lambda}
\sum _{b\in B_\Lambda} J_b \delta_{\triangle_k, X} s^D_b 
[ \prod_{l=1}^k g(X_l) - \prod_{l=1}^k g'(X_l) ]
\prod_{l=1}^kI(b\in \partial X_l)\nonumber \\
&&-\sum_{k=1}^\infty \frac{1}{k!}  \sum_{X_1, \cdots, X_k \in S_\Lambda}
\sum _{b\in B_\Lambda}\epsilon_b^y \delta_{b \triangle X, \triangle_k}s^D_b 
[ \prod_{l=1}^k g(X_l) - \prod_{l=1}^k g'(X_l) ]
\prod_{l=1}^kI(b\in \partial X_l)
\Big| (\epsilon M)^{-w(X)}\nonumber \\
 &&\leq ({\rm I })+({\rm  II}),
\end{eqnarray}
where  each term in the last line is defined by
\begin{eqnarray}
({\rm  I})&:=&  \sup_{c \in B_\Lambda} \sum_{X\in S_\Lambda}I(c\in \partial X)\Big|\sum_{k=2}^\infty \frac{1}{k!}  \sum_{X_1, \cdots, X_k \in S_\Lambda}
\sum _{b\in B_\Lambda} J_b s^D_b \delta_{\triangle_k, X}  
[ \prod_{l=1}^k g(X_l) - \prod_{l=1}^k g'(X_l) ]
\prod_{l=1}^kI(b\in \partial X_l)
\Big| (\epsilon M)^{-w(X)}, \nonumber  \\
({\rm  II})&:=&\sup_{c \in B_\Lambda}\sum_{X\in S_\Lambda}I(c\in \partial X)\Big| \sum_{k=1}^\infty \frac{1}{k!}  \sum_{X_1, \cdots, X_k \in S_\Lambda}
\sum _{b\in B_\Lambda} \epsilon_b^y s^D_b\delta_{b \triangle 
\triangle_k, X}
[ \prod_{l=1}^k g(X_l) - \prod_{l=1}^k g'(X_l) ]
\prod_{l=1}^kI(b\in \partial X_l)
 \Big|(\epsilon M)^{-w(X)}. \nonumber
\end{eqnarray}
Let us evaluate each term. 
An upper bound on (I) is
\begin{eqnarray}
({\rm  I})&\leq&  \sup_{c \in B_\Lambda} \sum_{k=2}^\infty \frac{1}{k!} \hspace{-2mm}\sum_{X_1, \cdots, X_k \in S_\Lambda} \hspace{-5mm} I(c\in \partial \triangle_k )
\Big|\sum _{b\in B_\Lambda} J_b  s^D_b
[ \prod_{l=1}^k g(X_l) - \prod_{l=1}^k g'(X_l) ]
\prod_{l=1}^kI(b\in \partial X_l)\Big|
(\epsilon M)^{-w(\triangle_k)} \nonumber  \\
 &\leq&  \sup_{c \in B_\Lambda}\sum_{k=2}^\infty \frac{1}{k!} \hspace{-2mm}\sum_{X_1, \cdots, X_k \in S_\Lambda} \hspace{-1mm}\sum_{l=1}^k I(c\in \partial X_l)
\Big|\sum _{b\in B_\Lambda} J_b  s^D_b  \prod_{l=1}^kI(b\in \partial X_l)\Big|
| \prod_{l=1}^k g(X_l) - \prod_{l=1}^k g'(X_l) |
(\epsilon M)^{-w(X_l)}
  \nonumber  \\
   &=&  \sup_{c \in B_\Lambda}\sum_{k=2}^\infty \frac{1}{(k-1)!} \hspace{-2mm}\sum_{X_1, \cdots, X_k \in S_\Lambda} \hspace{-5mm}I(c\in \partial X_1)
\Big|\sum _{b\in B_\Lambda} J_b  s^D_b\prod_{l=1}^kI(b\in \partial X_l)\Big|
| \prod_{l=1}^k g(X_l) - \prod_{l=1}^k g'(X_l) |
(\epsilon M)^{-w(X_l)},
 \nonumber  
\end{eqnarray}
where $w(\triangle_k) = \sum_{l=1}^k w(X_l)$,  
$I(c\in \partial \triangle_k)  \leq  \sum_{ l=1}^k  I ( c \in  \partial X_l) $ and permutation symmetry in the summation over $X_1, \cdots, X_k$ 
have been used. 
An inequality (\ref{tri})
enables us to evaluate  (I) as follows: 
\begin{eqnarray}
({\rm I}) &&\leq 
\sup_{c \in B_\Lambda}\sum_{k=2}^\infty \frac{1}{(k-1)!}
 \sum_{X_1, \cdots, X_k\in S_\Lambda} I(c\in \partial X_1)
 \Big|\sum_{b\in B_\Lambda} J_b s^D_b\prod_{l=1}^k  I(b\in \partial X_l)\Big|\nonumber\\
&&\quad\times\left[\sum_{l=1} ^k \prod_{j=1}^{l-1} |g(X_j)|(\epsilon M)^{-w(X_j)}
|g(X_l)-g'(X_l) | (\epsilon M)^{-w(X_l)}
 \prod_{j=l+1} ^k
|g'(X_j)| (\epsilon M)^{-w(X_j)}\right]
\nonumber\\
&&=
\sum_{k=2}^\infty \frac{1}{(k-1)!}
\sup_{c_1, \cdots, c_k \in B_\Lambda} \sum_{X_1, \cdots, X_k\in S_\Lambda}
 \Big|\sum_{b \in \partial X_1} J_b s^D_b\Big|
 \sum_{l=1} ^k \prod_{j=1}^{l-1} |g(X_j)|I(c_j\in \partial X_j)(\epsilon M)^{-w(X_j)}\nonumber\\
&&\quad\times
|g(X_l)-g'(X_l)|(\epsilon M)^{-w(X_l)}I(c_l\in \partial X_l)(\epsilon M)^{-w(X_j)}
\prod_{j=l+1} ^k
|g'(X_j)|I(c_j\in \partial X_j)(\epsilon M)^{-w(X_j)}.
\end{eqnarray}
(I) is bounded in terms of norms of $g,g'$.
\begin{eqnarray}
 (\rm I) &&\leq 
\|g-g'\|\sum_{k=2}^\infty \frac{1}{(k-1)!} \hspace{-1mm}
 \sum_{l=1} ^k \prod_{j=1}^{l-1}2 \| g \| /\Delta\prod_{j=l+1} ^k 2\| g'\|/\Delta
 = \|g-g'\|\sum_{k=2}^\infty \frac{1}{(k-1)!} \sum_{l=1} ^k (2\| g \|/\Delta)^{l-1} (2\|g'\|/\Delta)^{k-l} \nonumber \\
&&\leq \|g-g'\| \sum_{k=2} ^\infty \frac{k (\delta/\Delta)^{k-1}}{(k-1)!} = [e^{2\delta/\Delta} (1+2\delta/\Delta ) -1] \| g-g'\|,
\nonumber
\end{eqnarray}
where  $\Delta$ defined by (\ref{K2}) has been used.
Note that  the following inequality for any $g$
$$
\sup_{c\in B_\Lambda} \sum_{X \in S_\Lambda}I(c\in \partial X)|g(X)| (\epsilon M)^{-{w(X)}} 
\leq  \frac{2\| g\|}{\Delta}\leq  \frac{2\delta}{\Delta},
$$
has been used.
An upper bound on (II) is evaluated in the following
\begin{eqnarray}
({\rm  II})&:=&\sup_{c \in B_\Lambda}\sum_{X\in S_\Lambda}I(c\in \partial X)\Big| \sum_{k=1}^\infty \frac{1}{k!} \hspace{-2mm}\sum_{X_1, \cdots, X_k \in S_\Lambda}
\sum _{b\in B_\Lambda} \epsilon_b^ys^D_b \delta_{b \triangle 
\triangle_k, X}
[ \prod_{l=1}^k g(X_l) - \prod_{l=1}^k g'(X_l) ]
\prod_{l=1}^kI(b\in \partial X_l)
(\epsilon M)^{-w(X)}
\Big|\nonumber \\
&\leq&\sup_{c \in B_\Lambda}\sum_{X\in S_\Lambda}I(c\in \partial X) \sum_{k=1}^\infty \frac{1}{k!}\Big| \hspace{-2mm} \sum_{X_1, \cdots, X_k \in S_\Lambda}
\sum _{b\in B_\Lambda} \epsilon_b^ys^D_b \delta_{b \triangle \triangle_k, X} 
[ \prod_{l=1}^k g(X_l) - \prod_{l=1}^k g'(X_l) ]\Big|
\prod_{l=1}^kI(b\in \partial X_l)
(\epsilon M)^{-w(X)}.
\nonumber
\end{eqnarray}
Summation over $X$ in each  $k$-th term gives 
\begin{eqnarray}
({\rm  II})
&\leq&\sup_{c \in B_\Lambda} \sum_{k=1}^\infty \frac{1}{k!} \hspace{-2mm}\sum_{X_1, \cdots, X_k \in S_\Lambda}
\sum _{b\in B_\Lambda}| \epsilon_b^y| I(c\in \partial (b \triangle \triangle_k ) )   | \prod_{l=1}^k g(X_l) - \prod_{l=1}^k g'(X_l) |\prod_{l=1}^kI(b\in \partial X_l) (\epsilon M)^{-w(b \triangle \triangle_k )}\nonumber \\
&\leq &\sup_{c \in B_\Lambda} \sum_{k=1}^\infty \frac{1}{k!} \hspace{-2mm}\sum_{X_1, \cdots, X_k \in S_\Lambda}
\sum _{b\in B_\Lambda} |\epsilon_b^y| [ I(c\in \partial b )+ \sum_{l=1}^k    I(c \in \partial X_l)] 
|\prod_{l=1}^k g(X_l) - \prod_{l=1}^k g'(X_l) |
\prod_{l=1}^kI(b\in \partial X_l)(\epsilon M)^{-w(b \triangle \triangle_k )}
\nonumber \\
&=&\sup_{c \in B_\Lambda} \sum_{k=1}^\infty \frac{1}{k!} \hspace{-2mm}\sum_{X_1, \cdots, X_k \in S_\Lambda}
\sum _{b\in B_\Lambda} |\epsilon_b^y| [ I(b\in \partial c )+k    I(c \in \partial X_1)]
|\prod_{l=1}^k g(X_l) - \prod_{l=1}^k g'(X_l) |
\prod_{l=1}^kI(b\in \partial X_l)(\epsilon M)^{-\sum_{j=1}^k w(X_j)-1} \nonumber \\
&=& ({\rm II_A}) + ({\rm II_B}).
\end{eqnarray}
An explicit  decomposition
\begin{equation}
 \partial c := \bigsqcup_{a=1} ^{2(2d-1)} \{c_a \in B_\Lambda\},  \label{decomp}
 \end{equation}
 enables us to evaluate 
 the first term $(\rm II_A)$  in the last line, 
\begin{eqnarray}
&& ({\rm II_A}) 
 :=\sup_{c \in B_\Lambda} \sum_{k=1}^\infty \frac{1}{k!} \hspace{-2mm}\sum_{X_1, \cdots, X_k \in S_\Lambda}
\sum _{b\in B_\Lambda} |\epsilon_b^y|  I(b\in \partial c )
|\prod_{l=1}^k g(X_l) - \prod_{l=1}^k g'(X_l) |
\prod_{l=1}^kI(b\in \partial X_l) (\epsilon M)^{-\sum_{j=1}^k w(X_j) -1}\nonumber \\
&&\leq \sup_{c \in B_\Lambda}
\sum_{b\in \partial c} |\epsilon_b^y | \sum_{k=1}^\infty \frac{1}{k!}
\hspace{-2mm}
\sum_{X_1, \cdots, X_k\in S_\Lambda}\prod_{l=1}^kI(b\in \partial X_l)
\sum_{l=1} ^k \prod_{j=1}^{l-1} |g(X_j)|
|g(X_l)-g'(X_l) |
\prod_{j=l+1} ^k
|g'(X_j)|(\epsilon M)^{-\sum_{j=1}^k w(X_j) -1}
\nonumber\\
&&\leq  \sum_{k=1}^\infty \frac{1}{k!}
\sum_{l=1} ^k \prod_{j=1}^{l-1} \sup_{b_j \in B_\Lambda}\sum_{ X_j\in S_\Lambda} |g(X_j)| I(b_j\in \partial X_j)
(\epsilon M)^{-w(X_j) }
 \nonumber\\&&\quad\times
 \hspace{-1mm}\sup_{c \in B_\Lambda}\sum_{ X_l\in S_\Lambda} 
\sum_{b\in \partial c \cap \partial X_l } |\epsilon_b^y | 
|g(X_l)-g'(X_l) |(\epsilon M)^{-w(X_l) -1}
\prod_{j=l+1} ^k
\sup_{b_j \in B_\Lambda}\sum_{ X_j\in S_\Lambda} |g'(X_j)| I(b_j\in \partial X_j)(\epsilon M)^{-w(X_j) }
\nonumber\\
&&\leq  \sum_{k=1}^\infty \frac{1}{k!}
\sum_{l=1} ^k (\epsilon M)^{-1}\prod_{j=1}^{l-1} \sup_{b_j \in B_\Lambda}\sum_{ X_j\in S_\Lambda} |g(X_j)| I(b_j\in \partial X_j)
(\epsilon M)^{-w(X_j) }
 \nonumber\\&&\quad\times  \hspace{-2mm}
 \sum_{a=1} ^{2(2d-1)}  \hspace{-2mm} \sup_{c_a \in B_\Lambda} \sum_{ X_l\in S_\Lambda} 
\sum_{b\in \partial X_l } |\epsilon_b^y |I(c_a \in \partial X_l)
|g(X_l)-g'(X_l) |(\epsilon M)^{-w(X_l)}
\hspace{-2mm} 
  \prod_{j=l+1} ^k
\sup_{b_j \in B_\Lambda}\sum_{ X_j\in S_\Lambda} |g'(X_j)| I(b_j\in \partial X_j)
(\epsilon M)^{-w(X_j) }, \nonumber
\end{eqnarray}
$(\rm II_A)$ is bounded in terms of norms of $g,g'$.
\begin{eqnarray}
({\rm II_A}) &&\leq  (4d-2)(\epsilon M)^{-1} \epsilon_y/\Delta  \|g-g'\| \sum_{k=1}^\infty \frac{1}{k!} \sum_{l=1} ^k \Big( \frac{2\| g\|}{\Delta} \Big)^{l-1}
\Big( \frac{2\| g'\|}{\Delta} \Big)^{k-l}
\nonumber\\
&&\leq (4d-2)  (\epsilon M)^{-1}\epsilon_y/\Delta \|g-g'\| \sum_{k=1}^\infty \frac{1}{(k-1)!} \Big( \frac{2\delta}{\Delta} \Big)^{k-1}
 = (4d-2)(\epsilon M)^{-1} \epsilon_y/\Delta e^{2\delta/\Delta}  \|g-g'\|,
\end{eqnarray}
where   $\epsilon_y$ is defined by   (\ref{K2}).
The second term is evaluated as follows:
\begin{eqnarray}
&& ({\rm II_B}) 
 :=\sup_{c \in B_\Lambda} \sum_{k=1}^\infty \frac{1}{(k-1)!} \hspace{-2mm}\sum_{X_1, \cdots, X_k \in S_\Lambda}
\sum _{b\in B_\Lambda} |\epsilon_b^y|  I(c\in \partial X_1)(\epsilon M)^{-\sum_{j=1}^k w(X_j) -1}
|\prod_{l=1}^k g(X_l) - \prod_{l=1}^k g'(X_l) |
\prod_{l=1}^kI(b\in \partial X_l) \nonumber \\
&&\leq 
\sum_{k=1}^\infty \frac{1}{(k-1)!} \hspace{-2mm}\sup_{c_1, \cdots, c_k \in B_\Lambda} \sum_{X_1, \cdots, X_k \in S_\Lambda}
\sum _{b\in B_\Lambda} |\epsilon_b^y|  I(c_1\in \partial X_1)(\epsilon M)^{-\sum_{j=1}^k w(X_j) -1}
|\prod_{l=1}^k g(X_l) - \prod_{l=1}^k g'(X_l) |
\prod_{l=1}^kI(c_l \in \partial X_l) \nonumber \\
&&\leq (\epsilon M)^{-1 }  \epsilon_y/\Delta  \|g-g'\| \sum_{k=1}^\infty \frac{1}{(k-1)!} \sum_{l=1} ^k \Big( \frac{2\| g\|}{\Delta} \Big)^{l-1}
\Big( \frac{2\| g'\|}{\Delta} \Big)^{k-l}
\leq (\epsilon M)^{-1 } \epsilon_y /\Delta \|g-g'\| \sum_{k=1}^\infty \frac{k}{(k-1)!} \Big( \frac{2\delta}{\Delta} \Big)^{k-1}
\nonumber \\&& 
=(\epsilon M)^{-1 } \epsilon_y/\Delta  e^{2\delta/\Delta}(2\delta/\Delta+1) \|g-g'\|, \nonumber
\end{eqnarray}
where the inequality (\ref{K2}) has been used. 
These imply that an upper bound on (II) is given by
\begin{eqnarray}
({\rm II}) = ({\rm II_A})+ ({\rm II_B}) \leq(\epsilon M)^{-1 }  \epsilon_y/\Delta  e^{2\delta/\Delta}(2\delta/\Delta+4d-1) \| g-g'\|.
\end{eqnarray}
Therefore,  evaluations of (I) and (II) give the following  upper bound on the norm 
\begin{equation}
\|F(g)-F(g')\| \leq ({\rm I})+ ({\rm II})  \leq [e^{2\delta/\Delta}(2\delta/\Delta+1)-1+(\epsilon M)^{-1 } \epsilon_y/\Delta e^{2\delta/\Delta}(2\delta/\Delta+4d-1)] \| g-g'\|.
\end{equation}
Assume 
\begin{eqnarray}
 e^{2\delta/\Delta}(2\delta/\Delta+1)-1+(\epsilon M)^{-1 } \epsilon_y /\Delta e^{2\delta/\Delta}(2\delta/\Delta+4d-1 )\leq \frac{1}{2}.
\label{1/2}
\end{eqnarray}
To obtain the bound on $\| F(g) \|$,  let us evaluate $\| F(0)\|$ first. Since
$$
F(0) (X)= \frac{\sum_{b\in B_\Lambda}(\epsilon_b^y s^D_b-\epsilon_b^x )\delta _{X, b} }{ \sum_{b \in \partial X} J_b s^D_b },
$$
and the definition $\epsilon:=\sup_{c\in B_\Lambda}\sum_{b \in \partial c} 
 |\epsilon_b^y s^D_b-\epsilon_b^x  |,$ the norm of $F(0)$ is given by
\begin{eqnarray}
\| F(0)\| &=&  \sup_{c\in B_\Lambda}\sum_{X \in S_\Lambda} I(c\in \partial X)\Big|\sum_{b\in B_\Lambda}(\epsilon_b^y s^D_b-\epsilon_b^x )\delta _{X, b}  \Big| (\epsilon M)^{-w(X) }
 \nonumber \\
 &=&   \sup_{c\in B_\Lambda}\sum_{b \in B_\Lambda} I(c\in \partial b)|\epsilon_b^y s^D_b-\epsilon_b^x  | (\epsilon M)^{-1 }
\nonumber  \\
&=& \sup_{c\in B_\Lambda}\sum_{b \in \partial c} 
 |\epsilon_b^y s^D_b-\epsilon_b^x  | (\epsilon M)^{-1 } =M^{-1}
 .
\end{eqnarray}
 If $ M^{-1 } \leq \delta /2 ,$ then $\| F(g) \| \leq \delta$ is proven as follows:
$$
\| F(g) \| = \| F(g) -F(0) +F(0)\| \leq \| F(g) -F(0)\|  +\| F(0)\| \leq \frac{\| g \|}{2} +  \frac{\delta}{2}\leq \delta,
$$
The energy gap between $E_\pm$ is given by
\begin{eqnarray}E_+-E_- =&& -\sum_{k=2}^\infty \frac{1}{k!}  \sum_{ X_1, \cdots, X_k\in S_\Lambda }\sum _{c\in B_\Lambda} J_c 
\delta_{\triangle_k, \Lambda_L} s_c^D \prod_{l=1}^k g(X_l)I(c\in \partial X_l) \nonumber \\
  &&+\sum_{k=1}^\infty \frac{1}{k!}  \sum_{ X_1, \cdots, X_k\in S_\Lambda }
\sum _{c\in B_\Lambda} \epsilon_c^y \delta_{c\triangle  \triangle_k,\Lambda_L}s_c^D \prod_{l=1}^k g(X_l)I(c\in \partial X_l), 
\label{EG}\end{eqnarray}
An upper bound on $|E_+-E_-|$ is evaluated as 
\begin{eqnarray}
|E_+-E_- | \leq 
&& \Big| \sum_{k=2}^\infty \frac{1}{k!}  \sum_{ X_1, \cdots, X_k\in S_\Lambda }\sum _{c\in B_\Lambda} J_c 
\delta_{\triangle_k, \Lambda_L} s_c^D \prod_{l=1}^k g(X_l)I(c\in \partial X_l) \nonumber \\
  &&-\sum_{k=1}^\infty \frac{1}{k!}  \sum_{ X_1, \cdots, X_k\in S_\Lambda } \sum _{c\in B_\Lambda} \epsilon_c^y
  \delta_{c\triangle  \triangle_k,\Lambda_L}s_c^D \prod_{l=1}^k g(X_l)I(c\in \partial X_l) \Big|\nonumber \\
   \leq 
 && \sum_{k=2}^\infty \frac{1}{k!}  \sum_{ X_1, \cdots, X_k\in S_\Lambda }\Big| \sum _{c\in B_\Lambda} J_c  s_c^D \Big| 
\delta_{\triangle_k, \Lambda_L} \prod_{l=1}^k |g(X_l)|I(c\in \partial X_l) \nonumber \\
  &&+\sum_{k=1}^\infty \frac{1}{k!}  \sum_{ X_1, \cdots, X_k\in S_\Lambda } 
 \Big| \sum _{c\in B_\Lambda} \epsilon_c^y s_c^D\Big| \delta_{c\triangle  \triangle_k,\Lambda_L} \prod_{l=1}^k |g(X_l)|I(c\in \partial X_l)
 \nonumber \\
  \leq 
&& \sum_{k=2}^\infty \frac{1}{k!}  \sum_{ X_1, \cdots, X_k\in S_\Lambda }\Big| \sum _{c\in B_\Lambda} J_c  s_c^D \Big| 
\delta_{\triangle_k, \Lambda_L} (\epsilon M)^{\sum_{l=1} ^k w(X_l)}  \prod_{l=1}^k |g(X_l)| (\epsilon M)^{-w(X_l)} I(c\in \partial X_l) \nonumber \\
  &&+\sum_{k=1}^\infty \frac{1}{k!}  \sum_{ X_1, \cdots, X_k\in S_\Lambda } 
 \sum _{c\in B_\Lambda} |\epsilon_c^y | \delta_{c\triangle  \triangle_k,\Lambda_L} 
 (\epsilon M)^{\sum_{l=1} ^k w(X_l)} \prod_{l=1}^k |g(X_l)| (\epsilon M)^{-w(X_l)}I(c\in \partial X_l).
  \nonumber 
  \end{eqnarray}
The inequality  $\sum_{l=1}^kw(X_k) \geq  |\Lambda_L|$ for $\triangle_k = \Lambda_L $  gives
  \begin{eqnarray} 
|E_+-E_- | 
 \leq 
&& \sum_{k=2}^\infty \frac{1}{k!}  \sum_{ X_1, \cdots, X_k\in S_\Lambda }\Big| \sum _{c\in B_\Lambda} J_c  s_c^D \Big| 
\delta_{\triangle_k, \Lambda_L} (\epsilon M)^{|\Lambda_L|}  \prod_{l=1}^k |g(X_l)| (\epsilon M)^{-w(X_l)} I(c\in \partial X_l) \nonumber \\
  &&+\sum_{k=1}^\infty \frac{1}{k!}  \sum_{ X_1, \cdots, X_k\in S_\Lambda } 
 \sum _{c\in B_\Lambda} |\epsilon_c^y | \delta_{c\triangle  \triangle_k,\Lambda_L} 
 (\epsilon M)^{|\Lambda_L|} \prod_{l=1}^k |g(X_l)| (\epsilon M)^{-w(X_l)}I(c\in \partial X_l)
   \nonumber \\
 \leq 
&& \Big(\Big| \sum _{c\in B_\Lambda} J_c  s_c^D \Big| + \sum _{c\in B_\Lambda} |\epsilon_c^y | \Big) (\epsilon M)^{|\Lambda_L|} \sum_{k=2}^\infty \frac{1}{k!}  \prod_{l=1}^k \sup_{c_l \in B_\Lambda}
 \sum_{ X_l\in S_\Lambda } |g(X_l)| (\epsilon M)^{-w(X_l)} I(c\in \partial X_l) \nonumber \\
 \leq  &&
  \Big(\Big| \sum _{c\in B_\Lambda} J_c  s_c^D \Big| + \sum _{c\in B_\Lambda} |\epsilon_c^y | \Big) (\epsilon M)^{|\Lambda_L|} \sum_{k=2}^\infty \frac{1}{k!} \Big(\frac{\| g\|}{\Delta}\Big)^k \nonumber \\
 \leq  &&  \Big(\Big| \sum _{c\in B_\Lambda} J_c  s_c^D \Big| + \sum _{c\in B_\Lambda} |\epsilon_c^y | \Big) (\epsilon M)^{|\Lambda_L|} \exp^{(2)}  \Big(\frac{\delta}{\Delta}\Big).
\label{BEG}
\end{eqnarray}
Therefore, 
$$
|E_+-E_-| \leq A \Big(\Big| \sum _{c\in B_\Lambda} J_c  s_c^D \Big| + \sum _{c\in B_\Lambda} |\epsilon_c^y | \Big) 
(\epsilon M)^{|\Lambda_L|},
$$
where $A := \exp^{(2)} \alpha$
This completes the proof. $\Box$\\

\paragraph{Proof of  Theorem \ref{T3}} Lemma \ref{L6B} and the contraction mapping theorem
enable us to prove that the fixed point equation  $F(g) =g$
 has the unique solution $g$ for sufficiently weak perturbation of the XY interactions, under the  condition in Lemma \ref{L6B}.
 This solution gives energy eigenstates 
  \begin{equation}
 | \pm  \rangle =2^{-(|\Lambda_L|+1)/2}  \sum_{\sigma\in \{1,-1\}^{\Lambda_L}} \sigma_D (1\pm\sigma_{\Lambda_L})\exp \Big[-\frac{1}{2}\sum_{X\in S_\Lambda} g(X)  \sigma_X \Big] |\sigma\rangle, 
 \label{ketpm2}
 \end{equation} 
 corresponding to $|D\rangle \pm |D^c\rangle$ in the unperturbed  model.
  The condition in Lemma \ref{L6B} gives the bound on the energy gap
  $$
|E_+-E_-| \leq A \Big(\Big| \sum _{c\in B_\Lambda} J_c  s_c^D \Big| + \sum _{c\in B_\Lambda} |\epsilon_c^y | \Big) 
(2\epsilon/\delta)^{|\Lambda_L|}.
$$
If $\epsilon < \delta/2$, this energy gap becomes exponentially small  in the system size $|\Lambda_L|$.  This completes the proof. 
$\Box$

\subsection{Expansion for energy gap} 
Let  $C, D\in S_\Lambda$  be two different sub-lattices which
define sequences  of eigenvalues $s^C, s^D\in \{1,-1\} ^{\Lambda_L }$ in the unperturbed model $\bm \epsilon =\bm 0$.
Lemma \ref{L4} guarantees that there is no degeneracy
$H_\Lambda(s^C, \bm J, \bm 0) \neq H_\Lambda(s^D, \bm J, \bm 0)$  in the unperturbed model .
Here we evaluate energy gap between corresponding two energy eigenstates $|\pm\rangle'$ and $|\pm \rangle$ in perturbed model.
These  obey the following eigenvalue equations 
$$
-\sum_{b\in B_\Lambda}( J_b \sigma_b^x +\epsilon^x_b\sigma_b^z+\epsilon_b^y\sigma_b^y)| \pm \rangle = E_\pm |\pm\rangle,$$
$$
-\sum_{b\in B_\Lambda}( J_b \sigma_b^x +\epsilon^x_b\sigma_b^z+\epsilon_b^y\sigma_b^y) |\pm \rangle' =E_\pm' |\pm \rangle.
$$
Lemma \ref{L6B} guarantees  that there exists a function $\psi_\pm: \{1,-1\}^{\Lambda_L} \to {\mathbb R}$, such that
$$|\pm \rangle=2^{-(|\Lambda_L|+1)/2}\sum_{\sigma \in \{1,-1\}^{\Lambda_L}} \sigma_D(1\pm \sigma_{\Lambda_L}) \psi_\pm(\sigma) | \sigma \rangle.$$
Here, we show that  a  real valued function $\phi_\pm : \{1,-1\}^{\Lambda_L} \to {\mathbb  R}$ exists uniquely and 
the state $|\pm\rangle'$ can be represented in 
$$|\pm\rangle' =2^{-(|\Lambda_L|+1)/2} \sum_{\sigma \in \{1,-1\}^{\Lambda_L'}} \sigma_D(1\pm \sigma_{\Lambda_L})  \psi_\pm(\sigma) \phi_\pm(\sigma) | \sigma\rangle.$$
\begin{equation}
(1\pm \sigma_{\Lambda_L}) \Big( \sum _{b\in B_\Lambda}[ J_b\sigma_D^{(b)}  \psi_\pm(\sigma^{(b)})  \phi_\pm(\sigma^{(b)}) + \epsilon_b^x \sigma_b \sigma_D \psi_\pm(\sigma)  \phi_\pm(\sigma)-
\epsilon_b^y \sigma_b \sigma_D^{(b)}  \psi_\pm(\sigma^{(b)})  \phi_\pm(\sigma^{(b)})]
 + E_\pm ' \sigma_D \psi_\pm(\sigma) \phi_\pm(\sigma) \Big)=0.
\label{eigeneqpsi}
\end{equation}
The following relation
$$
\frac{\sigma^{(b)}_D}{\sigma_D} \frac{\psi_\pm(\sigma^{(b)})}{\psi_\pm(\sigma)}  =s_b^D(1\pm \sigma_{\Lambda_L})   \exp \Big[ \sum_{X \in S_\Lambda}  I(b \in \partial X)g(X) \sigma_X  \Big]
$$
and the eigenvalue equation (\ref{eigeneqpsi}) give 
\begin{equation}
(1\pm \sigma_{\Lambda_L}) \Big( \sum _{b\in B_\Lambda}  \Big[ (J_b-\epsilon_b^y\sigma_b)s_b^D \phi_\pm(\sigma^{(b)})   \exp \Big[\sum_{X \in S_\Lambda}  I(b \in \partial X)g(X) \sigma_X \Big] + \epsilon_b^x\sigma_b\phi_\pm(\sigma)\Big]  + E_\pm'  
 \phi_\pm(\sigma) \Big) =0.
\end{equation}
The eigenvalue equation for the reference state times $\phi(\sigma)$ is 
\begin{equation}
(1\pm \sigma_{\Lambda_L}) \Big(\sum _{b\in B_\Lambda} \Big[ (J_b-\epsilon_b^y\sigma_b)s_b^D \phi(\sigma)   \exp \Big[\sum_{X  \in S_\Lambda}  I(b \in \partial X)g(X) \sigma_X \Big] +\epsilon_b^x\sigma_b\phi(\sigma)\Big] + E_\pm  \phi_\pm(\sigma)\Big)=0.
\end{equation}
The difference between above two equations gives 
\begin{equation}
(1\pm \sigma_{\Lambda_L}) \Big(\sum _{b\in B_\Lambda} (J_b-\epsilon_b^y\sigma_b)s_b^D[ \phi_\pm(\sigma^{(b)})  -\phi_\pm(\sigma)] \exp \Big[\sum_{X  \in S_\Lambda}  I(b \in \partial X)g(X) \sigma_X \Big]  + (E_\pm' -E_\pm) \phi_\pm(\sigma) \Big) =0.
\end{equation}
To obtain the Kirkwood-Thomas equation for the state $|\pm\rangle'$, represent the function $\phi(\sigma)$ in terms of 
 a real valued  function $f(X)$ of an arbitrary subset $X  \in S_\Lambda $,
\begin{equation}
\phi_\pm(\sigma) =\frac{1}{2}(1\pm \sigma_{\Lambda_L}) \sum_{X  \in S_\Lambda} f(X) \sigma_X.
\end{equation}
This gives 
$$
\phi_\pm(\sigma^{(b)})  -\phi_\pm(\sigma) = -2 (1\pm \sigma_{\Lambda_L}) \sum_{X \in S_\Lambda} f(X) \sigma_X I(b \in \partial X)
$$
Then we have
\begin{eqnarray}
&& (1\pm \sigma_{\Lambda_L})\Big[ 2 \sum _{b\in B_\Lambda} (J_b-\epsilon_b^y\sigma_b)s_b^D  \sum_{Y \in S_\Lambda}  I(b \in \partial Y) f(Y)  \sigma_Y
 \exp \Big[ \sum_{X \in S_\Lambda}  I(b \in \partial X)g(X) \sigma_X \Big]   \nonumber \\
&&- (E_\pm'-E_\pm) \sum_{X \in S_\Lambda} f(X)\sigma_X \Big] = 0. \nonumber
\end{eqnarray}
Therefore
\begin{eqnarray}
&&(1\pm \sigma_{\Lambda_L})\Big[ \sum_{Y  \in S_\Lambda}  \Delta_Y f(Y)  \sigma_Y +2\sum _{b\in B_\Lambda} J_b s_b^D  \sum_{Y \in S_\Lambda}  I(b\in \partial Y) f(Y)  \sigma_Y\exp^{(1)} \Big[ \sum_{X  \in S_\Lambda}  I(b \in \partial X)g(X) \sigma_X \Big]  \\
 &&-2\sum _{b\in B_\Lambda} \epsilon_b^y\sigma_b s_b^D  \sum_{Y \in S_\Lambda}  I(b\in \partial Y) f(Y)  \sigma_Y\exp\Big[ \sum_{X  \in S_\Lambda}  I(b \in \partial X)g(X) \sigma_X \Big] -
 (E_\pm'-E_\pm) \sum_{X \in S_\Lambda} f(X)\sigma_X\Big]=0,  \nonumber 
\end{eqnarray}
where an energy gap $\Delta_Y$  for $Y\in S_\Lambda$ is defined by
$$
\Delta _Y:=  2 \sum_{b \in \partial Y} J_bs_b^D.
$$
The orthonormalization property (\ref{on})  gives
\begin{eqnarray}
 && \sum_{Y  \in S_\Lambda} 2\sum _{b\in \partial Y} f(Y)\sum_{k=1}^\infty \frac{1}{k!}  \sum_{ X_1, \cdots, X_k \in S_\Lambda}
J_bs_b^D  \delta_{\triangle_k\triangle  Y\triangle Z, \phi} \prod_{l=1}^k g(X_l) I(b\in \partial X_l)  
\nonumber   \\&&- \sum_{Y  \in S_\Lambda} 2\sum _{b\in \partial Y} f(Y)\sum_{k=0}^\infty \frac{1}{k!}  \sum_{ X_1, \cdots, X_k \in S_\Lambda}
\epsilon_b^y s_b^D  \delta_{\triangle_k\triangle  Y\triangle Z \triangle b, \phi}\prod_{l=1}^k g(X_l) I(b\in \partial X_l)  \nonumber  \\
&&\pm \sum_{Y  \in S_\Lambda} 2\sum _{b\in \partial Y} f(Y)\sum_{k=1}^\infty \frac{1}{k!}  \sum_{ X_1, \cdots, X_k \in S_\Lambda}
J_bs_b^D  \delta_{\triangle_k\triangle  Y\triangle Z, \Lambda_L} \prod_{l=1}^k g(X_l) I(b\in \partial X_l)  
  \\&&\mp  \sum_{Y  \in S_\Lambda} 2\sum _{b\in \partial Y} f(Y)\sum_{k=0}^\infty \frac{1}{k!}  \sum_{ X_1, \cdots, X_k \in S_\Lambda}
\epsilon_b^y s_b^D  \delta_{\triangle_k\triangle  Y\triangle Z \triangle b, \Lambda_L}\prod_{l=1}^k g(X_l) I(b\in \partial X_l)  
= (E_\pm'-E_\pm-\Delta_Z) [f(Z) \pm f(Z^c)] . \nonumber 
\end{eqnarray}
\begin{eqnarray}
 && \sum_{Y  \in S_\Lambda} 2\sum _{b\in \partial Y}[ f(Y)\pm f(Y^c)]\sum_{k=1}^\infty \frac{1}{k!}  \sum_{ X_1, \cdots, X_k \in S_\Lambda}
J_bs_b^D  \delta_{\triangle_k\triangle  Y\triangle Z, \phi} \prod_{l=1}^k g(X_l) I(b\in \partial X_l)  
  \\&&- \sum_{Y  \in S_\Lambda} 2\sum _{b\in \partial Y} [f(Y)\pm f(Y^c)]\sum_{k=0}^\infty \frac{1}{k!}  \sum_{ X_1, \cdots, X_k \in S_\Lambda}
\epsilon_b^y s_b^D  \delta_{\triangle_k\triangle  Y\triangle Z \triangle b, \phi}\prod_{l=1}^k g(X_l) I(b\in \partial X_l)   \nonumber \\
&&= (E_\pm'-E_\pm-\Delta_Z) [f(Z) \pm f(Z^c)] . \nonumber 
\end{eqnarray}
Define $e_\pm(C):=E_\pm'-E_\pm-\Delta_C$ and $e_\pm(X):= [f(X)\pm f(X^c)]/[f(C)\pm f(C^c)]$. 
for  $X\neq C.$\\

 For $Z=C$, 
 \begin{eqnarray}
 &&e_\pm(C)=   \sum_{Y \in S_\Lambda} 2\sum _{b\in \partial Y} J_bs_b^D e_\pm(Y)\sum_{k=1}^\infty \frac{1}{k!}  \sum_{ X_1, \cdots, X_k \in S_\Lambda}
  \delta_{\triangle_k, Y\triangle D}  \prod_{l=1}^k g(X_l) I(b\in \partial X_l) \nonumber \\
&&-  \sum_{Y \in S_\Lambda} 2\sum _{b\in \partial Y}\epsilon_b^y s_b^D e_\pm(Y)\sum_{k=0}^\infty \frac{1}{k!}  \sum_{ X_1, \cdots, X_k \in S_\Lambda}
 \delta_{\triangle_k, Y\triangle D \triangle b} \prod_{l=1}^k g(X_l) I(b\in \partial X_l) =: F(e)(C),
  \nonumber 
\end{eqnarray}
\begin{eqnarray}
 && e_\pm(Z)=\frac{1}{\Delta_Z-\Delta_C} \Big[ e_\pm(C)  e_\pm(Z)-  \sum_{Y \in S_\Lambda} 
 2\sum _{b\in \partial Y} J_b s_b^D  e_\pm(Y)\sum_{k=1}^\infty \frac{1}{k!}  \sum_{ X_1, \cdots, X_k \in S_\Lambda}
\delta_{\triangle_k, Y\triangle Z}\prod_{l=1}^k g(X_l) I(b\in \partial X_l)  \nonumber  \\
 &&+  \sum_{Y \in  S_\Lambda} 2\sum _{b\in \partial Y} \epsilon_b^y s_b^De_\pm(Y)\sum_{k=0}^\infty \frac{1}{k!}  \sum_{ X_1, \cdots, X_k \in S_\Lambda}
 \delta_{\triangle_k, Y\triangle Z \triangle b} \prod_{l=1}^k g(X_l) I(b\in \partial X_l)  \Big]=:F(e_\pm)(Z).
\end{eqnarray}
These two equations defines a fixed point equation $F(e)=e$, whose  solution $e$ gives the state $|\pm\rangle'$ except its
normalization. To prove the uniqueness of the solution, define a norm for the function $e$ by 
\begin{equation}
\| e_\pm \| :=|e_\pm(C)|+ \sum_{X \in S_\Lambda} |\Delta_X-\Delta_C| |e_\pm(X)|.
\label{norm}
\end{equation}
The following theorem implies that  there is no level crossing between two arbitrary states $|\pm\rangle', |\pm\rangle $   against a sufficiently weak perturbation of  the XY-exchange interactions, if we impose a condition to break the ${\mathbb Z}_2$ symmetry. \\

 {\theorem \label{T3.2} Consider the random bond Heisenberg XYZ model defined by the Hamiltonian (\ref{XYZ})
  For two different sub-lattices $C, D \in S_\Lambda$, 
let $s^C, s^D \in \{1,-1\}^{\Lambda_L}$ be their corresponding sequences defined by (\ref{sD}). 
If the sequence of XY-exchange $\bm \epsilon$ is sufficiently small, there exists a sufficiently small constant $\delta>0$ depending on the sequence of coupling constants $(\bm J, \bm \epsilon)$, such that 
the energy gap $E_\pm'-E_\pm$ in the perturbed model satisfies
 $$H_\Lambda^{\rm XYZ}(s^C, \bm J, \bm 0) - H_\Lambda^{\rm XYZ}(s^D, \bm J, \bm 0)-\delta  < E_\pm' -E_\pm <  H_\Lambda^{\rm XYZ}(s^C, \bm J, \bm 0) - H_\Lambda^{\rm XYZ}(s^D, \bm J, \bm 0)+\delta,$$
 for almost all $\bm J \in {\mathbb R}^{B_\Lambda}$.
Theorem \ref{T3.2}  implies that 
the sign of $E_\pm'-E_\pm$ is identical to that of  $ H_\Lambda(s^C, \bm J, \bm 0) - H_\Lambda(s^D, \bm J, \bm 0)$ 
 for $\delta <  |\ H_\Lambda(s^C, \bm J, \bm 0) - H_\Lambda(s^D, \bm J, \bm 0)|$.}\\

The following lemma and the contraction mapping theorem are helpful to prove Theorem \ref{T3.2}.\\

\noindent
{\lemma \label{L7}
Consider the random bond Heisenberg XYZ model  under the condition in Lemma \ref{L6}, namely 
\begin{eqnarray}
\frac{3}{2}\geq e^{2\delta/\Delta}(2\delta/\Delta+1)+ \epsilon_y /\Delta e^{2\delta/\Delta}(2\delta/\Delta+4d-1 ).
\end{eqnarray}
If 
$
K:=2 (1+|\Delta_C|/\Delta')(e^{2\delta/\Delta}-1)+2 \epsilon_y'/\Delta' e^{2\delta/\Delta}+\delta/\Delta'  <1
$
  for 
\begin{equation}
\Delta' := \inf_{Y\in S_\Lambda} |\Delta_Y-\Delta_C|, \ \ \  \epsilon_y':=\Delta'\sup_{c\in B_\Lambda,Y\in S_\Lambda} I(c\in \partial Y)\frac{ 2 \sum _{b\in \partial Y} |\epsilon_b^y| }{|\Delta_Y -\Delta_C|},
\label{Delta'}
\end{equation}
then for almost all $\bm J \in { \mathbb R}^{B_\Lambda},$  the following norms are  bounded by
\begin{equation}
\|F(e_\pm)-F(e_\pm') \| \leq  K \| e_\pm-e_\pm' \|,   \ \ \|F(e_\pm) \| \leq \delta, \ \   {\rm for}\  \| e_\pm\|, \| e_\pm'\| \leq \delta,
\end{equation} 
}

\noindent
{\bf Proof.}  For lighter notation, we remove indices $\pm$ from $e_\pm$. 
The difference between  two evaluations of  energy gap  is
\begin{eqnarray}
 &&|F(e)(C)-F(e')(C)| = \Big| \sum_{Y \in  S_\Lambda} 2\sum _{b\in \partial Y} J_bs_b^D [e(Y)-e'(Y)]\sum_{k=1}^\infty \frac{1}{k!}  \sum_{ X_1, \cdots, X_k \in S_\Lambda}
  \delta_{\triangle_k, Y\triangle D}  \prod_{l=1}^k g(X_l) I(b\in \partial X_l) \nonumber \\
  &&-  \sum_{Y \in S_\Lambda} 2\sum _{b\in \partial Y}\epsilon_b^y s_b^D [e(Y)-e'(Y)]\sum_{k=0}^\infty \frac{1}{k!}  \sum_{ X_1, \cdots, X_k \in S_\Lambda}
 \delta_{\triangle_k, Y\triangle D \triangle b} \prod_{l=1}^k g(X_l) I(b\in \partial X_l) \Big|.
   \end{eqnarray}
The triangle inequality gives \begin{eqnarray}
&&|F(e)(C)-F(e')(C)| \leq \sum_{Y \in S_\Lambda} \Big| 2\sum _{b\in \partial Y} J_b s_b^D \Big| |e(Y)-e'(Y)|\sum_{k=1}^\infty \frac{1}{k!}  \sup_{b_1, \cdots, b_k \in B_\Lambda} \sum_{ X_1, \cdots, X_k \in S_\Lambda}
\delta_{\triangle_k, Y\triangle D}  \prod_{l=1}^k g(X_l) I(b_l\in \partial X_l)\Big| \nonumber \\
 &&+  \sum_{Y \in S_\Lambda} 2 \sum _{b\in \partial Y} |\epsilon_b^y | |e(Y)-e'(Y)|\sum_{k=0}^\infty \frac{1}{k!}  \sum_{ X_1, \cdots, X_k \in S_\Lambda}
 \delta_{\triangle_k, Y\triangle D \triangle b}  \prod_{l=1}^k g(X_l) I(b\in \partial X_l)\Big| \nonumber \\
 &&\leq  \sum_{Y \in S_\Lambda}| \Delta_Y ||e(Y)-e'(Y)|\sum_{k=1}^\infty \frac{1}{k!}  
 \prod_{l=1}^k \sup_{b_l \in B_\Lambda} \sum_{  X_l \in S_\Lambda} |g(X_l)| I(b_l \in \partial X_l)  \nonumber  \\
 &&+\sum_{Y \in S_\Lambda} 2 \sum _{b\in \partial Y}  |\epsilon_b^y | |e(Y)-e'(Y)|\sum_{k=0}^\infty \frac{1}{k!}  
 \prod_{l=1}^k \sup_{b_l \in B_\Lambda} \sum_{  X_l \in S_\Lambda} |g(X_l)| I(b_l \in \partial X_l) \nonumber \\
  && \leq  
   \sum_{Y \in S_\Lambda} | \Delta_Y| |e(Y)-e'(Y)|\sum_{k=1}^\infty \frac{1}{k!} \Big(\frac{\delta}{\Delta}\Big)^k+2\sum_{b\in \partial D}| \epsilon_b^y | 
  \sum_{Y\in S_\Lambda}
   |e(Y)  -e'(Y) |\sum_{k=0}^\infty \frac{1}{k!} \Big(\frac{\delta}{\Delta}\Big)^k
  \nonumber \\
  && \leq  \| e-e'\|\Big[   (1+|\Delta_C|/\Delta') (e^{2\delta/\Delta} -1) + \epsilon_y'/\Delta'e^{2\delta/\Delta}\Big],
  \end{eqnarray}
  where  $\Delta'$ and $\epsilon_y'$ are defined by (\ref{Delta'}). 
The norm between $F(g)$ and $F(g')$ is 
\begin{eqnarray}
 && \|F(e)-F(e')\| = |F(e)(C)-F(e')(C)|+
\sum_{Z\in S_\Lambda} \Big| e(C)  e(Z) -e'(C) e'(Z) \nonumber \\
 &&- \sum_{Y\in S_\Lambda} 2\sum _{b\in \partial Y} J_b s_b^D [ e(Y)-e'(Y)]\sum_{k=1}^\infty \frac{1}{k!}  \sum_{ X_1, \cdots, X_k \in S_\Lambda}
\delta_{
X_1 \triangle \cdots \triangle X_k, Y\triangle Z} \prod_{l=1}^k g(X_l) I(b\in \partial X_l)  
  \nonumber \\
   &&+  \sum_{Y \in S_\Lambda} 2\sum _{b\in \partial Y} \epsilon_b^y s_b^D [ e(Y)-e'(Y)]\sum_{k=0}^\infty \frac{1}{k!}  \sum_{ X_1, \cdots, X_k \in S_\Lambda}
\delta_{X_1 \triangle \cdots \triangle X_k, Y\triangle Z \triangle b} \prod_{l=1}^k g(X_l) I(b\in \partial X_l)  \Big|
  \nonumber \\
 &&\leq
  |F(e)(C)-F(e')(C)|+ ({\rm I}) +({\rm II}) + ({\rm III}),
 \end{eqnarray}
  where  each term in the last line is defined by
 \begin{eqnarray} 
&&({\rm I}) := 
 \sum_{Z\in S_\Lambda}  | e(C)  e(Z) -e'(C) e'(Z)|  \nonumber \\
 &&({\rm II}):= \sum_{Y,Z\in S_\Lambda} \Big| 2\sum _{b\in  \partial Y} J_b s_b^D  [e(Y)-e'(Y)]\sum_{k=1}^\infty \frac{1}{k!}  \sum_{ X_1, \cdots, X_k \in S_\Lambda}
\delta_{
X_1 \triangle \cdots \triangle X_k, Y\triangle Z} \prod_{l=1}^k g(X_l) I(b\in \partial X_l)  \Big|
  \nonumber \\
   &&({\rm III}):= \sum_{Y,Z\in S_\Lambda} \Big| 2\sum _{b\in  \partial Y} \epsilon_b^y s_b^C  [ e(Y)-e'(Y)]\sum_{k=0}^\infty \frac{1}{k!}  \sum_{ X_1, \cdots, X_k \in S_\Lambda}
\delta_{X_1 \triangle \cdots \triangle X_k, Y\triangle Z \triangle b} \prod_{l=1}^k g(X_l) I(b\in \partial X_l)  \Big|. \nonumber 
 \end{eqnarray}
Each can be  bounded as follows:   
 \begin{eqnarray} 
&&({\rm I}) \leq  | e(C)-e'(C)|\sum_{Z\in S_\Lambda} |e(Z)|  + |e(C)|\sum_{Z\in S_\Lambda} |e(Z)-e'(Z)|
 \nonumber \\
 &&= | e(C)-e'(C)| \sum_{Z\in S_\Lambda} |\Delta_Z- \Delta_C| |e(Z)|/\Delta' + |e(C)|/\Delta'  \sum_{Z\in S_\Lambda} |\Delta_Z- \Delta_C| |e(Z)-e'(Z)| \nonumber \\
 &&\leq \| e \| /\Delta' [ |e(C)-e'(C)|  +  \sum_{Z\in S_\Lambda} |\Delta_Z- \Delta_C| |e(Z)-e'(Z)|] \leq 
  \delta/\Delta' \|e-e'\|,
 \end{eqnarray}
  \begin{eqnarray}
 &&({\rm II})= \sum_{Y\in S_\Lambda} \Big| 2\sum _{b\in  \partial Y} J_b s_b^D \Big| |e(Y)-e'(Y)|\sum_{k=1}^\infty \frac{1}{k!}  \sup_{_1, \cdots, c_k \in B_\Lambda}\sum_{ X_1, \cdots, X_k \in S_\Lambda}
 \prod_{l=1}^k| g(X_l)| I(c_l\in \partial X_l)  
  \nonumber \\
  &&\leq  \sum_{Y\in S_\Lambda} | \Delta_Y| |e(Y)-e'(Y)|\sum_{k=1}^\infty \frac{1}{k!}  
\prod_{l=1}^k\sup_{c_l \in B_\Lambda} \sum_{  X_l\in S_\Lambda} |g(X_l)| I(c_l\in \partial X_l)  \nonumber \\
&&\leq (1+|\Delta_C|/\Delta')(e^{2\delta/\Delta}-1)\| e-e'\|,
\end{eqnarray}
\begin{eqnarray}
&&({\rm III}) \leq  \sum_{Y\in S_\Lambda} 2\sum _{b\in  \partial Y} |\epsilon_b^y |  | e(Y)-e'(Y)|\sum_{k=0}^\infty \frac{1}{k!}  \sum_{ X_1, \cdots, X_k \in S_\Lambda}
 \prod_{l=1}^k |g(X_l)| I(b\in \partial X_l)    \nonumber \\
  &&\leq  \sum_{Y \in S_\Lambda}   2\sum _{b\in  \partial Y} |\epsilon_b^y |  |e(Y)-e'(Y)|\sum_{k=0}^\infty \frac{1}{k!}  
\prod_{l=1}^k\sup_{c_l \in B_\Lambda} \sum_{  X_l\in S_\Lambda} |g(X_l)| I(c_l\in \partial X_l)  
  \nonumber \\
  &&\leq \epsilon_y'/\Delta'e^{2\delta/\Delta}\| e-e'\|.
\end{eqnarray}
Therefore
\begin{eqnarray}
 &&\|F(e)-F(e')\| \leq   |F(e)(C)-F(e')(C)|+ ({\rm I}) +({\rm II}) + ({\rm III}) \\
   &&\leq \| e-e'\|\Big[   (1+|\Delta_C|/\Delta') (e^{2\delta/\Delta} -1) + \epsilon_y'/\Delta'e^{2\delta/\Delta} \Big]+\delta/\Delta' \|e-e'\| \nonumber \\&&
 + (1+|\Delta_C|/\Delta')(e^{2\delta/\Delta}-1)\| e-e'\|+\epsilon_y'/\Delta' e^{2\delta/\Delta}\| e-e'\|
  \nonumber \\
 && =[2 (1+|\Delta_C|/\Delta')(e^{2\delta/\Delta}-1)+2 \epsilon_y'/\Delta' e^{2\delta/\Delta}+\delta/\Delta' ]\| e-e'\| = K \| e-e'\| .
\end{eqnarray}
If $K < 1$, $G$ is a contraction mapping.
  This completes the proof of Lemma \ref{L7}. $\Box$

\paragraph{Proof of  Theorem \ref{T3.2}} 
Lemma \ref{L7} and  the contraction mapping theorem guarantee the unique solution $e_\pm(C)$ of the  fixed point equation 
under the condition on the perturbation of the sequence of coupling constants in Lemma \ref{L7}. Then, there  exists $\delta>0$, such that
$$|E_\pm'-E_\pm- \Delta_C| = |e_\pm(C)|\leq  \delta. $$  This completes the proof of  Theorem \ref{T3.2}. $\Box$\\

\subsection{Remarks on results for the random bond Heisenberg XYZ model}
Theorem \ref{T3} claims that  an  arbitrary  energy gap $|E_+-E_-|$ 
between split energy eigenvalues  is exponentially small in the system size $|\Lambda_L|$, also in the random bond Heisenberg XYZ 
model, as in  the transverse field EA model with weak transverse fields.   
 A spontaneous symmetry  breaking of the $\mathbb Z_2$ symmetry is expected  also in this model.  
 In our expansion method,  however, we cannot give  any results in the infinite-volume limit.\\

\noindent
{\bf Acknowledgment}    \\
It is pleasure to thank  
Chokri Manai and Simone Warzel  for helpful discussions in degeneracy of energy eigenvalues. We are grateful to Sei Suzuki for a helpful comment on quantum annealing.  
C.I. is supported by JSPS (21K03393).


\end{document}